\documentclass[11pt]{article}
\usepackage[a4paper,top=2cm,left=2.5cm,right=2.5cm,bottom=2cm]{geometry}
\usepackage[dvipsnames]{xcolor}
\usepackage{amssymb,amsfonts,amsmath,amsthm}
\usepackage{verbatim}
\usepackage{cite}
\usepackage{microtype}
\newcommand{\be}{\begin{equation}}
\newcommand{\ee}{\end{equation}}
\newcommand{\dd}{\mathrm{d}}
\newcommand{\Rl}{{R}}
\newcommand{\omegal}{{\omega}}

\newcommand{\bl}{{b}}
\newcommand{\epsc}{\epsilon^{(0)}}
\newcommand\adsS[2]{AdS${}_{#1} \times $ S${}^{#2}$}
\definecolor{vub}{RGB}{0,52,154}
\usepackage[
colorlinks=true, % false: boxed links; true: colored links
linkcolor=vub,   % color of internal links
citecolor=vub,   % color of links to bibliography
filecolor=magenta, % color of file links
urlcolor=cyan,
]
{hyperref}
\allowdisplaybreaks
\usepackage{setspace}
\numberwithin{equation}{section}

\begin{document}

\begin{center}
	\baselineskip=16pt
	
	\vspace{-5cm}
	  
	\hfill IFT-UAM/CSIC-25-26  
	
	\vskip 0.5cm

	{\LARGE \bf  
	Supersymmetric solutions of non-relativistic\\[10pt] 11-dimensional supergravity 
	}
	\vskip 2em
	{\large \bf  Chris D. A. Blair}
	\vskip 0.6em
	{\it  
	Instituto de Física Teórica UAM/CSIC, \\ %C/ Nicolás Cabrera 13-15,
	Universidad Autónoma de Madrid, Cantoblanco, Madrid 28049, Spain \\  {\tt c.blair@csic.es}}
	\vskip 0.5cm 
\end{center}

\begin{abstract}
Eleven-dimensional supergravity has a non-relativistic variant obtained by taking a limit associated with the M2 brane.
Consistency of this non-relativistic supergravity requires constraints.
There is one choice of constraints which keeps the maximal amount of supersymmetry transformations, and another which only keeps half.
I discuss supersymmetric solutions of this theory, based on limits of the M2 and M5 solutions.
These limits involve either a \emph{scaling} of the number of branes or a \emph{smearing} in certain directions, and have been argued to produce non-Lorentzian bulk geometries appearing in novel versions of the AdS/CFT correspondence.
I show that the scaled solutions are solutions of the maximally supersymmetric version of the non-relativistic supergravity, while the smeared solutions only give solutions of the half-maximally supersymmetric version. I show these solutions are all supersymmetric by solving the Killing spinor equations, and discuss usual and unusual forms of supersymmetry enhancement. I also discuss a simple supersymmetric AdS${}_3$ background, and point out that the BTZ black hole is a solution of non-relativistic 11-dimensional supergravity.
\end{abstract} 

\tableofcontents

\section{Introduction}

Eleven-dimensional supergravity \cite{Cremmer:1978km} and its solutions cut a window onto M-theory.
The M2 and M5 branes are revealed as supersymmetric geometries solving the equations of motion and admitting half the maximal amount of Killing spinors.
The near-horizon limits of these solutions lead to the maximally supersymmetric \adsS{4}{7} and \adsS{7}{4} geometries, pointing the way to dual holographic descriptions of the M2 and M5 field theories.

In this paper, I will discuss what happens to these solutions, and their supersymmetry, when taking the membrane non-relativistic limit of 11-dimensional supergravity \cite{Blair:2021waq, Bergshoeff:2024nin}.
This limit renders the geometry of spacetime non-relativistic (or non-Lorentzian), taking us from the metric description to a degenerate structure involving three `longitudinal' and eight `transverse' vielbeins, linked to the dimensionality of the membrane worldvolume.

I am motivated by broader interests in such limits, which can be formulated as BPS decoupling limits associated with each supersymmetric extended object (string or brane) in 10- and 11-dimensions \cite{Gomis:2000bd, Danielsson:2000gi,Blair:2023noj}.
Many familiar scenarios in string theory are in fact controlled by such limits, such as the low energy limit on a D-brane worldvolume and the relationship between M-theory and Matrix Theory on D0-branes. 
Indeed, AdS/CFT itself is another example, where the decoupling limit in question manifests in the bulk as the usual near-horizon limit and non-Lorentzian geometry appears only asymptotically \cite{Blair:2024aqz}.

In curved spacetime, these limits rely on the introduction of new (string or $p$-brane) versions of non-relativistic Newton-Cartan geometry, building on \cite{Andringa:2012uz,Harmark:2017rpg, Bergshoeff:2018yvt}.
A proliferating duality web of examples is now available \cite{Blair:2023noj,Gomis:2023eav}.
One of the ambitions here is the pursuit of new corners of holography in string theory, involving non-Lorentzian bulk geometries.
There has been recent progress in formulating such geometries by engineering non-relativistic limits of (near-horizon) supergravity brane solutions \cite{Lambert:2024uue, Lambert:2024yjk,Fontanella:2024rvn,Fontanella:2024kyl, Blair:2024aqz,Lambert:2024ncn,Harmark:2025ikv,Guijosa:2025mwh,Blair:2025prd}. The holographic dual can be argued for by working out the accompanying decoupling limit applied to the appropriate worldvolume field theory.

It is rather easier to take limits of solutions directly than to deal with the limit of the theories that they solve.
A priori, it is not guaranteed that the limits of supergravity solutions are solutions of the limits of supergravity, and if so, that they are still supersymmetric.
Examples have been generated using various tricks, such as dualising along null directions, introducing extra scalings of parameters that are not part of the intrinsic definition of the limit, or smearing.
In this paper, I want to discuss (some of) these approaches, and to draw a link between them and the surprising structure of supersymmetry in the non-relativistic supergravity limit.

I will focus on 11-dimensional supergravity.
The famously unique 11-dimensional supergravity admits at least two non-relativistic variants, obtained by taking limits associated with the M2 
\cite{Blair:2021waq, Bergshoeff:2024nin} and M5 branes \cite{Bergshoeff:2025grj}.
The former case is the only example where the non-relativistic limit has been extended to include the fermionic sector of a maximal supergravity.\footnote{For a review of other non-relativistic supergravity limits, see \cite{Bergshoeff:2022iyb}. 
A string non-relativistic limit of half-maximal supergravity in ten dimensions was constructed in \cite{Bergshoeff:2021tfn} and supersymmetric brane solutions thereof investigated in \cite{Bergshoeff:2022pzk}.
As discussed in \cite{Bergshoeff:2025grj}, the relationship between the limit of the 11-dimensional supergravity and the half-maximal theory of \cite{Bergshoeff:2021tfn} is actually unclear, and will not be pursued further in this paper.
}
Consistency of supersymmetry in the limit requires the imposition of constraints on the bosonic geometry.
In \cite{Bergshoeff:2024nin}, two sets of consistent constraints were identified, leading to two versions of the resulting non-relativistic supergravity.
The first version admits maximal supersymmetry (32 independent spinorial transformation parameters), and the constraints involve a highly non-trivial tower of both bosonic and fermionic constraints, setting various torsions, curvatures or their derivatives to zero.
The second version keeps only half the supersymmetries (16 independent spinorial transformation parameters), and leads to a much reduced set of (purely bosonic) constraints.\footnote{Caveats apply to the analysis of \cite{Bergshoeff:2024nin}, which in particular relied to an extent on evidence based on linearised calculations. Here I will simply assume that the approach of \cite{Bergshoeff:2024nin} correctly determined the transformations and constraint structure of the non-relativistic supergravity. Another way of reading the present paper is as a further check that the results of \cite{Bergshoeff:2024nin} are sensible, by testing them on other explicit solutions.}

Meanwhile, limits of supergravity solutions have been separately considered.
In particular, in \cite{Lambert:2024uue} a limit of the M2 brane supergravity solution was taken, and argued to give a non-Lorentzian geometry dual to a related non-relativistic limit of the low-energy membrane worldvolume field theory.
The limit of this M2 solution required scaling -- by hand -- the constant in the harmonic function in order to obtain a non-trivial result.
This scaling can be viewed as a scaling of the number of branes sourcing the background (in this case it is a large $N$ rescaling).
The same mechanism has been demonstrated to engineer non-Lorentzian backgrounds, in a variety of limits associated with strings and branes in both 10- and 11-dimensions, in \cite{Avila:2023aey,Lambert:2024uue, Lambert:2024yjk,Fontanella:2024kyl,Fontanella:2024rvn, Blair:2024aqz}.

While the physical significance of this additional scaling remains puzzling, the resulting non-relativistic M2 solution of \cite{Lambert:2024uue} was shown in \cite{Bergshoeff:2024nin} not only to satisfy the equations of motion and constraints of the maximally supersymmetric version of non-relativistic 11-dimensional supergravity, but in its near-horizon limit to admit 32 Killing spinors.
This seems like a desirable property (and gave the authors of \cite{Bergshoeff:2024nin} no little confidence that the monstrous constraint structure they had uncovered was not totally vacuous).
Curiously though, treating the limit of the M2 background as an ansatz for a solution depending only on longitudinal coordinates, it was noticed that the equations of motion are in fact trivially satisfied, and this could be exploited to find there exists a family of backgrounds sharing the same supersymmetry as the M2 solution.

In addition, it has since been realised that there is an alternative route to a good non-Lorentzian limit of brane solutions, which does not introduce additional parameter scalings (or rather does so in a more `natural' fashion).
This route involves smearing the brane solution on the directions which are longitudinal to the limit \cite{Blair:2024aqz, Lambert:2024ncn,Harmark:2025ikv}.
The smeared backgrounds can also be argued to have the correct symmetry properties to be holographic duals of accompanying non-relativistic field theories \cite{Lambert:2024ncn}.
It is then natural to ask to what extent the properties of these backgrounds are consistent with them appearing as solutions, supersymmetric or not, of the underlying non-relativistic supergravity theories. 

\vspace{0.5em}
\noindent 
It follows that supersymmetric solutions of non-relativistic 11-dimensional supergravity warrant further attention.
In this paper, I will study the solutions that result from taking limits of the M2 and M5 supergravity solutions.
The results and structure of this paper are summarised below:

\vspace{0.5em}\noindent 
{\emph{Review of non-relativistic 11-dimensional supergravity.}
In section \ref{sec:setting}, I introduce the constituents of the theory obtained by taking the non-relativistic M2 limit of 11-dimensional supergravity.
This is based on a membrane Newton-Cartan (M2NC) geometry, 
and I will refer to the theory as M2NC SUGRA.
Similarly, I will refer to the limit used as the M2NC limit (elsewhere \cite{Blair:2023noj, Blair:2024aqz} I have used the terminology Non-Relativistic M-theory (NRMT) for the result of this limit).
In section \ref{sec:bosons}, I describe the geometry and give the bosonic equations of motion.
In section \ref{sec:fermions}, I introduce the maximally supersymmetric and half-maximally supersymmetric versions of M2NC SUGRA, their constraints, and discuss the Killing spinor equations including the possibility of solutions leading to infinite-dimensional enhancements of supersymmetry.
In section \ref{sec:solutions}, as a warm-up I discuss how longitudinal three-dimensional geometries with non-positive Ricci scalar provide simple solutions of the theory, including AdS${}_3$ and the BTZ black hole, and describe a solution of the Killing spinor equations for a longitudinal AdS${}_3$ background.

\vspace{0.5em}\noindent 
{\emph{Limits of the M2 solution.}
In section \ref{sec:M2}, I revisit the M2 supergravity solution and its M2NC limits.
Firstly, in section \ref{sec:M2scaled}, I consider the limit in which the number of branes of the solution is scaled by hand, leading to a solution localised in the longitudinal directions of the M2NC geometry, originally written down in \cite{Lambert:2024uue}.
It was shown in \cite{Bergshoeff:2024nin} that the near-horizon form of this solution admitted 32 Killing spinors in the maximally supersymmetric version of M2NC SUGRA.
I carefully redo this calculation, and show that in fact the near-horizon solution is part of a class of backgrounds admitting an infinite family of Killing spinor solutions.
Secondly, in section \ref{sec:M2smeared}, I consider the limit in which the M2 solution is smeared in the longitudinal directions of the M2NC limit.
This leads to an M2NC background localised in the transverse directions of the M2NC geometry.
I confirm that this is indeed a solution of the M2NC SUGRA, but now only in its half-maximal version.
I then show that it admits a Killing spinor solution involving 8 Killing spinors.
In fact, slightly more generally I show that this is the case for a solution characterised by a function which is harmonic in the doubly transverse directions, but can have an arbitrary dependence on the coordinates which are longitudinal in the M2NC geometry but transverse to the M2 brane.
Unlike for the scaled solution, I find no supersymmetry enhancement in the near-horizon limit.

\vspace{0.5em}\noindent 
\emph{Limits of the M5 solution.}
In section \ref{sec:M5}, I consider the M5 supergravity solution and its M2NC limits.
Firstly, in section \ref{sec:M5scaled}, I consider the limit in which the number of branes of the solution is scaled by hand, leading to a solution localised in the longitudinal directions of the M2NC geometry.
This solution obeys the constraints of the maximally supersymmetric version of M2NC SUGRA.
Like the M2, the longitudinal dependence of the solution is in fact not constrained by the equations of motion, but unlike for the M2, there is no family of supersymmetric solutions: the near-horizon form is singled out by the process of solving the Killing spinor equations. 
However, the Killing spinor equations also admit an infinite-dimensional set of solutions given in terms of arbitrary solutions of the (1+1)-dimensional Dirac equation, giving an enhancement of supersymmetry.
Secondly, in section \ref{sec:M5smeared}, I consider the limit in which the M5 of the solution is first smeared in the longitudinal directions of the M2NC limit.
This leads to an M2NC background localised in the transverse directions of the M2NC geometry, originally written down in \cite{Lambert:2024ncn}.
Just as for the M2, this is only a solution of the half-maximal version of M2NC SUGRA.
I find Killing spinor solutions again involving arbitrary solutions of the two-dimensional massless Dirac equation, with no further enhancement in the near-horizon limit. I discuss how this relates to the field theory expectations of \cite{Lambert:2024yjk}.
Lastly, in section \ref{sec:M5spherical}, I consider an alternative limit of the M5, where the transverse radial coordinate is treated as longitudinal in the M2NC limit.
This again leads to a supersymmetric solution of the maximally supersymmetric M2NC SUGRA.

\vspace{1em}
\noindent
Finally, I present a brief summary and conclusions in section \ref{sec:summary}.

\section{Non-relativistic 11-dimensional supergravity} 
\label{sec:setting}

\subsection{Geometry and equations of motion} 
\label{sec:bosons}

I review here the formulation of the 11-dimensional membrane Newton-Cartan supergravity (M2NC SUGRA) \cite{Blair:2021waq, Bergshoeff:2024nin}.
I start with the bosonic sector.
The bosonic fields of 11-dimensional supergravity are the metric and three-form, for which I introduce 
the following reparametrisation:
\be
G_{\mu\nu} = c^2 \tau_{\mu\nu} + c^{-1} E_{\mu\nu} \,,\quad
C_{\mu\nu\rho} = -c^3  \epsilon_{ABC} \tau^A{}_\mu \tau^B{}_\nu \tau^C{}_\rho + c_{\mu\nu\rho} \,,
\label{defmnc}
\ee
where I defined
\be
\tau_{\mu\nu} \equiv \tau^A{}_\mu \tau^B{}_\nu \eta_{AB} \,,\quad
E_{\mu\nu} \equiv e^a{}_\mu e^b{}_\nu \delta_{ab} \,,
\ee
with $A=0,1,2$ and $a=3,\dots,10$ labelling longitudinal and transverse tangent space indices, respectively.
Here $\eta_{AB}$ is the three-dimensional Minkowski metric with mostly plus signature.
The inverse of $G_{\mu\nu}$ can similarly be parametrised in terms of quantities $\tau^\mu{}_A$ and $e^\mu{}_a$ satisfying 
\be
\tau^A{}_\mu e^\mu{}_a = 0 \,\quad
e^a{}_\mu \tau^\mu{}_A = 0 \,,\quad
e^\mu{}_a e^b{}_\mu = \delta_a^b \,,\quad
\tau^\mu{}_A \tau^B{}_\mu = \delta_A^B\,,\quad
\tau^\mu{}_A \tau^A{}_\nu + e^\mu{}_a e^a{}_\nu = \delta^\mu_\nu \,.
\label{NCcomplete} 
\ee
I collectively refer to $\tau^A{}_\mu$, $e^a{}_\mu$ (and their inverses) as M2NC vielbeins.
They will be used to flatten curved indices. 

The M2NC limit is defined by sending $c \rightarrow \infty$ in \eqref{defmnc}. 
In taking this limit, an additional bosonic field is introduced, in order to realise a finite limit of the supergravity action.
This field will be non-zero in some of the solutions considered later, so I now explain how and why it is introduced.
Treating \eqref{defmnc} as a field redefinition and inserting these expressions into the action of 11-dimensional supergravity leads to an expansion of the form
\be
S    = - c^3\int d^{11} x \,  \Omega\,\tfrac{1}{4!}   f^{(-)}{}^{abcd}  f^{(-)}_{abcd} + S_0 + O(c^{-3}) \,.
\label{Sexpansion}
\ee
The precise details for the $c$-independent terms, $S_0$, can be found in \cite{Blair:2021waq, Bergshoeff:2024nin}.
The term proportional to $c^3$ involves the anti-self-dual part of the totally transverse projection of the field strength of the M2NC three-form potential, defined through:
\be
f^{(-)}_{abcd} = \tfrac12 ( f_{abcd} - \tfrac{1}{4!} \epsilon_{abcd}{}^{efgh} f_{efgh} ) \,,\quad
f_{abcd} \equiv e^\mu{}_a e^\nu{}_b e^\rho{}_c e^\sigma{}_d f_{\mu\nu\rho\sigma}\,,\quad
f_{\mu\nu\rho\sigma} = 4 \partial_{[\mu} c_{\nu\rho\sigma]} \,.
\ee
(I define $f_{Aabc}$, $f_{ABab}$, $f_{ABCa}$ by similar contractions with $e^\mu{}_a$ and $\tau^\mu{}_A$.)
Lastly, $\Omega$ is a measure factor, given by
\be
\Omega = - \tfrac{1}{3!8!} \epsilon^{\mu \nu\rho\sigma_1 \dots \sigma_8} \epsilon_{ABC} \epsilon_{a_1 \dots a_8} \tau^A{}_{\mu} \tau^B{}_{\nu} \tau^C{}_{\rho} e^{a_1}{}_{\sigma_1} \dots e^{a_8}{}_{\sigma_8} \,.
\ee
The $c \rightarrow \infty$ limit of \eqref{Sexpansion} should define the action for the M2NC supergravity.
This limit is obstructed by the term proportional to $c^3$.
To circumvent this obstruction, an auxiliary bosonic field $\lambda_{abcd}$ is introduced. This is taken to be anti-self-dual, obeying $\lambda_{abcd} = - \tfrac{1}{4!} \epsilon_{abcd}{}^{efgh} \lambda_{efgh}$.
Using this field, an equivalent action to \eqref{Sexpansion} is provided by
\be
S    =  S_0- \int d^{11} x \,\Omega \,\tfrac{2}{4!}  \lambda_{abcd}  f^{(-) abcd}  + c^{-3} \int d^{11} x \,\Omega\, \tfrac{1}{4!}  \lambda_{abcd} \lambda^{abcd} + O(c^{-3}) \,.
\label{Sexpansionprime}
\ee
Prior to taking the limit, the equation of motion of $\lambda_{abcd}$ sets $\lambda_{abcd} = c^3f^{(-)}_{abcd}$, and substituting this in \eqref{Sexpansionprime} gives \eqref{Sexpansion}.
After taking the limit, the first two terms of \eqref{Sexpansionprime} give the action of M2NC SUGRA, and here 
$\lambda_{abcd}$ becomes a Lagrange multiplier imposing that $f^{(-)}_{abcd}=0$.

Then $\tau^A{}_\mu$, $e^a{}_\mu$, $c_{\mu\nu\rho}$ and $\lambda_{abcd}$ form the bosonic sector of the M2NC SUGRA. 
\begin{subequations}\label{bosonictransf}%
In addition to usual diffeomorphisms, these fields transform under various local bosonic symmetries as follows:
\begin{align}
\delta \tau^A{}_\mu  &= \alpha \,\tau^A{}_\mu- \Lambda^A{}_B \tau^B{}_\mu \,, & %\delta \tau^\mu{}_A &= -\alpha \,\tau^\mu{}_A+\Lambda^B{}_A \tau^\mu{}_B  - \lambda_A{}^a e^\mu{}_a \,,
\\[4pt]
 \delta e^a{}_\mu  &= -\tfrac12 \alpha \,e^a{}_\mu- \Lambda^a{}_b e^b{}_\mu + \lambda_A{}^a \tau^A{}_\mu \,,&
%\delta e^\mu{}_a  &= \tfrac12 \alpha \,e^\mu{}_a+\Lambda^b{}_a e^\mu{}_b\,,
\\[4pt]
\delta c_{\mu\nu\rho} &= 3 \partial_{[\mu} \theta_{\nu\rho]}- 3 \,\epsilon_{ABC} \lambda^A{}_a e^a{}_{[\mu} \tau^B{}_{\nu} \tau^C{}_{\rho]} \,, \\[4pt]
\delta \lambda_{abcd} & = - \alpha\, \lambda_{abcd}+ 4 \Lambda^e{}_{[a} \lambda_{|e|bcd]}  +  \tfrac{1}{4!} P^{(-)}_{abcd}{}^{efgh} 4 \lambda^A{}_{[e} f_{|A|fgh]} \,,
\end{align}
\end{subequations}
with (infinitesimal) parameters $\Lambda^A{}_B$ for $\mathrm{SO}(1,2)$ rotations, $\Lambda^a{}_b$ for $\mathrm{SO}(8)$ rotations, $\lambda^A{}_a$ for Galilean membrane boosts mixing the longitudinal and transverse sectors, and $\alpha$ corresponding to an emergent dilatation symmetry. In addition $\theta_{\mu\nu}$ is the two-form gauge transformation parameter.
Here $P^{(-)}_{abcd}{}^{efgh}$ denotes an anti-self-dual projector.

To compactly write down the action, equations of motion and supersymmetry transformations requires introducing geometrical notions of connections and curvatures.
This is explained at length in \cite{Bergshoeff:2024nin}.
I collect here the minimal definitions to make this paper self-contained (in doing so I gloss over various subtleties important for the analysis of \cite{Bergshoeff:2024nin} but which can be ignored in the applications I consider in this paper).
Firstly, define the following:
\be
T_{\mu\nu}{}^A = 2 \partial_{[\mu} \tau^A{}_{\nu]} \,,\quad
\Omega_{\mu\nu}{}^a = 2 \partial_{[\mu} e^a{}_{\nu]} \,,
\label{defTOmega}
\ee
where $T_{\mu\nu}{}^A$ is conventionally called the torsion of the longitudinal vielbein, and I will refer to $\Omega_{\mu\nu}{}^a$ as the coefficients of anholonomy for the transverse vielbein.
\begin{subequations}\label{connections}
These appear, together with components of the field strength $f_{\mu\nu\rho\sigma}$, in the following connections associated with the bosonic transformations of \eqref{bosonictransf}:
\begin{align}
 \omegal_\mu{}^{AB} & =  T_{\mu}{}^{[AB]} - \tfrac12 \tau^C{}_\mu  T^{AB}{}_C \,,\\[4pt]
 \omegal_\mu{}^{ab} & =  \Omega_\mu{}^{[ab]} - \tfrac12 e_{c\mu}  \Omega^{abc} + \tfrac14 \tau_{A\mu} \epsilon^{ABC} {f}_{BC}{}^{ab}+\tfrac13 e^{[a}{}_\mu  T^{b] A}{}_A\,,
 \label{connections_nohats_ab}\\[4pt]
 \omegal_\mu{}^{Aa} & = \tfrac12  \Omega_\mu{}^{Aa} - \tfrac12 e_{\mu b}  \Omega^{Aab}
- \tfrac{1}{4} e^b{}{}_\mu { f}^a{}_{b BC} \epsilon^{BCA}
- \tau^A{}_\mu\tfrac13  { f}_{012}{}^a \,,\\[4pt]
 \bl_\mu & = \tfrac13 e^a{}_\mu  T_a{}^A{}_A \,.
\end{align}
\end{subequations}
Correspondingly, I define curvatures by: \begin{subequations}\label{curvatures}
\begin{align}
\Rl_{\mu\nu}{}^{AB} &= 2\,\partial_{[\mu}\omegal_{\nu]}{}^{AB} + 2\,\omegal_{[\mu}{}^{AC}\omegal_{\nu]C}{}^B{} \,,\\[4pt]
\Rl_{\mu\nu}{}^{ab} & = 2 \partial_{[\mu} \omegal_{\nu]}{}^{ab} + 2 \label{Rmunuab}
\omegal_{[\mu}{}^{ac} \omegal_{\nu]c}{}^b \,,
\\[4pt]
\Rl_{\mu\nu}{}^{Aa} &= 2\,\partial_{[\mu}\omegal_{\nu]}{}^{Aa} + 2\,\omegal_{[\mu}{}^{AB}\omegal_{\nu]B}{}^a + 2\,\omegal_{[\mu}{}^{ab}\omegal_{\nu]}{}^{A}{}_{b} + 3\,\bl_{[\mu}\omegal_{\nu]}{}^{Aa}\,,\\[4pt]
\Rl_{\mu\nu} & = 2 \partial_{[\mu} \bl_{\nu]} \,.
\end{align} 
\end{subequations} 
The equations of motion can be formulated in terms of these curvatures.
However, in general, the expressions \eqref{curvatures} do not give covariant objects, taking into account the transformations of the connections \eqref{connections} under the bosonic symmetries \eqref{bosonictransf}.
As explained in more detail in \cite{Bergshoeff:2024nin}, this requires introducing modified curvatures.
To formulate the equations of motion, the only definition that is needed is the following improved fully covariant $\mathrm{SO}(8)$ curvature:
\be
\label{improvedR}
\begin{split}
  \breve{R}_{\mu\nu}{}^{ab} &= \Rl_{\mu\nu}{}^{ab} 
 + 4\,\tau_{A[\mu}\omega_{\nu]B}{}^{[a}T^{b]\{AB\}} +\tau_{A[\mu}\epsilon^{ABC}\omega_{\nu]B}{}^cf_{Cc}{}^{ab}
   \\ & \quad  +2\,e_{c[\mu}\omega_{\nu]A}{}^{[a}T^{b]cA} - e_{c[\mu}\omega_{\nu]A}{}^cT^{abA} - \frac23\,e^{[a}{}_{[\mu}T^{b]cA}\omega_{\nu] Ac}\,.
\end{split}
\ee
Then, taking all possible longitudinal and transverse projections, I can write down the equations of motion of the vielbein fields:
\begin{subequations}\label{EOMCOV}%
\begin{align}
\breve{R}_{ab}{}^{ab}  & =  T^{a\{AB\}} T_{a\{AB\}} 
+ \tfrac{1}{36} f_{Aabc} f^{Aabc}
\label{tauEom1Cov} \,,\\[4pt]
\label{tauEom2Cov}
 0  &= e^{\mu a} D_{\mu} T_{a\{AB\}}  + \tfrac{1}{12} f_{\{A}{}^{abc} f_{B\} abc} \,,\\[4pt]
0 & = e^\nu{}_b  D_\nu T^{ab}{}_A + T_{b\{A B\}} T^{ab}{}^B  -\tfrac16 f^{abcd} f_{Abcd} + \tfrac12 \epsilon_{ABC} f^{abcB} T_{bc}{}^C  \,,
\label{tauEom3Cov}
\\[4pt]
0 & =
\breve{R}_{Aca}{}^c - \tfrac16 \lambda_{abcd} f_A{}^{bcd} + \tau^{\mu B} D_\mu T_{a\{AB\}} 
\,,
\label{ECovEom1}
\\[4pt]
0 \nonumber  &= 2 \breve{R}_{(a|c|b)}{}^c 
-\tfrac12 \epsilon_{ABC} T_{(a}{}^{cA} f_{b)c}{}^{BC} 
- 2 T_a{}^{\{BC\}} T_{b\{BC\} } 
\\[4pt]  & \qquad  - \tfrac12 f_{acdB} f_b{}^{cdB} - \tfrac13 \lambda_{acde} f_b{}^{cde} 
+ \tfrac{1}{18} \delta_{ab} f_{cdeA} f^{cdeA}\,,
\end{align} 
where the following covariant derivative is used:
\be
\begin{split}
D_\mu T_{a\{AB\}} & = \partial_\mu T_{a\{AB\}} - \omegal_\mu{}^b{}_a T_{b\{AB\}}- \omegal_\mu{}^C{}_{A} T_{a\{CB\}} - \omegal_\mu{}^C{}_{B} T_{a\{AC\}} 
\\ &\qquad + \omegal_{\mu \{A}{}^b T_{|ab|B\}} - \tfrac12 \bl_\mu T_{a\{AB\}} \,.
\end{split}
\ee
The equation of the motion of the three-form can be expressed explicitly as: 
\be
\begin{split} 
&
\partial_\sigma \left(
\Omega \left[
4 \tau^{[\mu}{}_A e^{\nu}{}_b e^\rho{}_c e^{\sigma]}{}_d f^{Abcd} 
- 6 \epsilon^{ABC} e^{[\mu}{}_a e^\nu{}_b \tau^\rho{}_A \tau^{\sigma]}{}_B T^{ab}{}_C 
+2 e^{[\mu}{}_a e^\nu{}_b e^\rho{}_c e^{\sigma]}{}_d \lambda^{abcd} 
\right] 
\right) 
\\ & + \tfrac12 \tfrac{1}{4!4!} \epsilon^{\mu\nu\rho \sigma_1 \dots \sigma_8} f_{\sigma_1 \dots \sigma_4} f_{\sigma_5 \dots \sigma_8} 
=0 \,,
\end{split}
\label{EomChere}
\ee
while as mentioned above the equation of motion of $\lambda_{abcd}$ sets $f^{(-)}_{abcd} = 0$.
All the preceding equations of motion follow from the variation of the action of the M2NC SUGRA.
There is also an equation of motion which does not follow from the action.
This is called the Poisson equation and it is given by:\footnote{Note that the relative sign of the $\lambda^2$ term was incorrect in the original versions of \cite{Bergshoeff:2024nin}.}
\be
\breve{R}_{Aa}{}^{Aa} + \tfrac{1}{4!} \lambda_{abcd} \lambda^{abcd}=0 \,,\quad
\breve{R}_{Aa}{}^{Ab}  \equiv  \Rl_{Aa}{}^{Ab} + \tfrac18\, \Rl_{AB}{}^{AB}\delta_a{}^b\,.
\label{Poisson}
\ee
This transforms into \eqref{ECovEom1} under boosts.
\end{subequations}

\subsection{Supersymmetry and constraints}
\label{sec:fermions} 

I now explain the realisation of supersymmetry in the M2NC SUGRA, as worked out in \cite{Bergshoeff:2024nin}.
The only fermionic field of 11-dimensional SUGRA is the gravitino, which I parametrise as:
\be
\Psi_\mu = c^{-1} \psi_{+\mu} + c^{1/2} \psi_{-\mu} \,, 
\label{ExpFer}
\ee
using the fact that the limit is intended to break 11-dimensional Lorentz invariance in order to define chiral spinors such that\footnote{My gamma matrices $(\gamma^A, \gamma^a)$ obey $\{\gamma^A, \gamma^B\} = 2 \eta^{AB}$, $\{\gamma^a, \gamma^b\} = 2 \delta^{ab}$, $\gamma^A \gamma^a = - \gamma^a\gamma^A$, and $\gamma^{012} \gamma^{3\dots10} = +1$.}
\be
\gamma_{012} \psi_{\pm \mu} = \pm \psi_{\pm \mu} \,.
\ee
The spinors $\psi_{\pm \mu}$ transform under the local bosonic symmetries as:
\begin{subequations}
\begin{align}
\label{bosonictransf_ferminus}
\delta \psi_{-\mu}& = + \tfrac12 \alpha \psi_{-\mu}- \tfrac14 \Lambda^A{}_B \gamma_A{}^B \psi_{-\mu} - \tfrac14 \Lambda^a{}_b \gamma_a{}^b \psi_{-\mu}\,,\\
\delta \psi_{+\mu}& = - \alpha \psi_{+\mu} - \tfrac14 \Lambda^A{}_B \gamma_A{}^B \psi_{+\mu} - \tfrac14 \Lambda^a{}_b \gamma_a{}^b \psi_{+\mu}- \tfrac12 \lambda_A{}^a \gamma^A{}_a \psi_{-\mu}  \,.
\label{bosonictransf_ferplus}
\end{align}
\end{subequations}
They also exhibit an emergent shift (or superconformal) symmetry which can be viewed as the superpartner of the bosonic dilatations (see below for the form of this transformation).

The realisation of local supersymmetry in the theory defined by the limit is a complicated affair.
Given a spinor $\epsilon$ appearing as a supersymmetry transformation parameter, it is similarly decomposed as $\epsilon = c^{-1} \epsilon_+ + c^{1/2} \epsilon_-$ with $\gamma_{012} \epsilon_\pm = \pm \epsilon_\pm$. 
Obstructions manifest in the form of divergences appearing in the limit of the supersymmetry transformations of the gravitinos, $\psi_{\pm \mu}$.
These divergences can be removed by imposing constraints.
The variations of these constraints lead to new constraints. 
It was argued in \cite{Bergshoeff:2024nin} that the following two options define consistent versions of 11-dimensional membrane Newton-Cartan supergravity, in which the tower of constraints terminates.

\vspace{1em}
\noindent $\bullet$ \emph{Maximal supersymmetry.}
In order to realise local supersymmetry with thirty-two independent parameters, $\epsilon_\pm$, the following bosonic constraints are sufficient:
\begin{subequations}\label{maxcons}%
\begin{align}
 T_{ab}{}^A & = 0 \,,\quad  T_a{}^{\{AB\}} = 0 \,,\quad  f_{Aabc} = 0 \,,\quad  f_{abcd} = 0 \,,
 \label{maxcons_torsion}
 \\[4pt]
\breve{R}_{ab}{}^{cd} &= 0 \,,\quad 
\breve{R}_{Aa}{}^{bc} = 0 \,,\quad 
D_a \breve{R}_{A(b}{}^A{}_{c)} = 0\,,\quad D_a \lambda_{bcde} = 0 \,,
\label{maxcons_curv}
\end{align} 
\end{subequations}
where the curvatures appearing are defined in \eqref{improvedR} and \eqref{Poisson} (in fact given \eqref{maxcons_torsion}, $\breve{R}_{\mu\nu}{}^{ab}$ of \eqref{improvedR} reduces to $R_{\mu\nu}{}^{ab}$ of \eqref{Rmunuab}).
The covariant derivative acting on $\breve{R}_{A}{}_{a}{}^A{}_b$ and $\lambda_{abcd}$ only involves the $\mathrm{SO}(8)$ and dilatation connections. The boost connection terms that would ordinarily also appear vanish owing to the fact that $\breve{R}_{A}{}_{a}{}^A{}_b$ and $\lambda_{abcd}$  boost into $\breve{R}_{Aa}{}^{bc}$ and $f_{Aabc}$, respectively, which are set to zero. 

These constraints imply that all the bosonic equations of motion following from the action are satisfied, that is, all the equations \eqref{EOMCOV} other than the Poisson equation.
There are also fermionic constraints, which will of course be identically satisfied as I consider purely bosonic solutions of the theory.

With the constraints \eqref{maxcons} imposed, the fermionic transformations of $\psi_{\pm \mu}$ are:
\begin{subequations}
\label{maxsusytransfs}
\begin{align}
\delta\psi_{-\mu} & = D_\mu \epsilon_- +  \tau^A{}_\mu \gamma_A \eta_-  \,, \\[4pt]
\delta \psi_{+\mu} & = D_\mu \epsilon_+  - \tfrac{1}{12}  \tau^A{}_\mu \gamma_A \tfrac{1}{4!} \lambda_{bcde} \gamma^{bcde}  \epsilon_+
 - \tfrac18 e^a{}_\mu  \tfrac{1}{4!} \lambda_{bcde} \gamma^{bcde} \gamma_a \epsilon_-  \notag \\ & \qquad  + \tau^A{}_\mu \rho_{+A}  - \tfrac12 e^a{}_\mu \gamma_a\eta_- \,,
\end{align}
\end{subequations}
\begin{subequations}
\label{defDeps}%
where the covariant derivatives of the supersymmetry parameters are: 
\begin{align} 
D_\mu \epsilon_- & = 
 (\partial_\mu + \tfrac14  \omega_{\mu}{}^{ab} \gamma_{ab} + \tfrac14  \omega_\mu{}^{AB}\gamma_{AB} - \tfrac12  b_\mu ) \epsilon_-\,,\\[4pt]
 D_\mu \epsilon_+ & = 
 (\partial_\mu  + \tfrac{1}{4} \omega_\mu{}^{ab} \gamma_{ab}
+ \tfrac{1}{4}  \omega_\mu{}^{AB} \gamma_{AB} 	+  b_\mu )\epsilon_+
+ \tfrac12  \omega_\mu{}^{Aa} \gamma_{Aa} \epsilon_-
 \,.
\end{align} 
Here $\eta_-$ and $\rho_{+A}$ are the parameters of the emergent fermionic shift symmetry; the latter is constrained to obey $\gamma^A \rho_{+A} = 0$. (The subscripts $\pm$ on these spinors reflect their eigenvalues under $\gamma_{012}$, just as for $\psi_{\pm \mu}$ and $\epsilon_{\pm}$.)
\end{subequations}

\vspace{1em}
\noindent $\bullet$ \emph{Half-maximal supersymmetry.}
Another option is to keep only half the independent supersymmetry transformations.
This can be done consistently by setting 
\begin{align}
 T_{ab}{}^A & = 0 \,,\quad    f_{abcd} = 0 \,,\quad \epsilon_- = 0  \,.
\label{halfcons}
\end{align} 
These constraints are \emph{not} sufficient to impose that the bosonic equations of motion following from the action are satisfied, and given a background obeying \eqref{halfcons} but not \eqref{maxcons}, these still need to be checked.

With the constraints \eqref{halfcons} imposed, the fermionic transformations of $\psi_{\pm \mu}$ are:
\begin{subequations}
\label{halfsusytransfs}
\begin{align}
 \delta \psi_{-\mu} & =  \tau_{A\mu} \left(
\tfrac12  T_{a}{}^{\{AB\}} \gamma_B \gamma^a
+\tfrac14 ( \eta^{AB} - \tfrac13 \gamma^A \gamma^B ) \tfrac{1}{3!} f_{Babc} \gamma^{abc} \right) \epsilon_+
+  \tau^A{}_\mu \gamma_A \eta_- 
\,,\\[4pt] \notag
\delta \psi_{+\mu} & = (\partial_\mu  + \tfrac{1}{4} \omega_\mu{}^{ab} \gamma_{ab}
+ \tfrac{1}{4}  \omega_\mu{}^{AB} \gamma_{AB} 	+  b_\mu )\epsilon_+
 - \tfrac{1}{12}  \tau^A{}_\mu \gamma_A  \tfrac{1}{4!} \lambda_{abcd} \gamma^{abcd} \epsilon_+
\\[4pt] &  \qquad  +\tfrac{1}{12} \tfrac{1}{6} e^a{}_\mu  f_{A bcd} ( \gamma_a{}^{bcd} - 6 \delta_a^b \gamma^{cd} ) \gamma^A \epsilon_+
 + \tau^A{}_\mu \rho_{+A}  - \tfrac12 e^a{}_\mu \gamma_a\eta_- 
 \,.
\end{align}
\end{subequations}
This version of supersymmetry is unconventional.
In particular, the supersymmetry algebra does not close into diffeomorphisms, as would ordinarily be expected.
On the other hand, the condition $\epsilon_- = 0$ is somewhat physically appealing as it seems to reflect that the M2NC limit itself could be viewed as arising from a 1/2 BPS brane.
In this paper, I will refer to this possibility as the half-maximally supersymmetric version of the M2NC SUGRA, although this might not be the optimal name given the non-standard nature of its `supersymmetry' transformations and the possibility of confusion with genuine half-maximal supergravities in lower dimensions such as \cite{Bergshoeff:2021tfn}.

\vspace{1em}
\noindent
The Killing spinor equations of the M2NC SUGRA are obtained by setting the above fermionic transformations, \eqref{maxsusytransfs} or \eqref{halfsusytransfs}, to zero.
Notice that these equations involve not only the supersymmetry transformation parameters $\epsilon_\pm$ but also the fermionic shift symmetry parameters $\eta_-$ and $\rho_{+A}$. 
Generically it is necessary that non-trivial solutions of the Killing spinor equations involve non-zero $\eta_-$ and $\rho_{+A}$.
Note though that, as shift symmetries, there are no solutions of the Killing spinor equations involving purely $\eta_-$ and $\rho_{+A}$: if $\epsilon_\pm = 0$, the only solutions to $\delta \psi_\pm = 0$ are $\eta_- = 0 = \rho_{+A}$. Therefore in any solution to the Killing spinor equations one can think of $\eta_-$ and $\rho_{+A}$ as being determined in terms of the genuine supersymmetry transformation parameters $\epsilon_\pm$.

One can view the presence of non-zero $\eta_-$ and $\rho_{+A}$ in a solution to the Killing spinor equations as necessary to preserve the condition $\psi_{\pm \mu} = 0$ as usual.
Alternatively, viewing the gravitino components which transform under the shift as unphysical degrees of freedom, the parameters $\eta_-$ and $\rho_{+A}$ appearing in the Killing spinor equation are merely there to absorb the part of the supersymmetry transformation acting on the unphysical components, regarding this as irrelevant.  

Regardless, the Killing spinor equations can always be treated by first solving for $\eta_-$ and $\rho_{+A}$ in terms of $\epsilon_\pm$ and their derivatives, so as to then backsubstitute and rewrite the Killing spinor equations purely in terms of $\epsilon_\pm$. 
For example, the longitudinal part of the equation $\delta \psi_{+\mu} = 0$ always determines $\rho_{+A}$, leading to:
\be
\rho_{+A} = - \tau^\mu{}_A D_\mu \epsilon_+ + \tfrac{1}{12} \gamma_A \tfrac{1}{4!} \lambda_{bcde} \gamma^{bcde} \epsilon_+ \,, 
\ee
subject to the consistency condition $\gamma^A \rho_{+A} = 0$.
This implies a sort of curved space massive Dirac equation for $\epsilon_+$.
However, it can also be convenient not to solve for $\eta_-$ and $\rho_{+A}$ a priori, in order to make a particular choice of these parameters which facilitates solving the full equations more straightforwardly.
A very important point though is that solving the Dirac equation following from $\gamma^A \rho_{+A} = 0$ is necessarily weaker than making a particular choice of $\rho_{+A}$ and solving the resulting individual equations $D_{A} \epsilon_+ = \dots$. If one only adopts the latter approach, it is likely that one may miss (infinitely many) solutions.

Indeed, an interesting complication which can arise when solving these Killing spinor equations is the appearance of (infinite-dimensional) supersymmetry enhancements, beyond what would be possible in the more familiar relativistic setting.

For instance, consider the flat M2NC geometry, with constant vielbeins, $\tau^A{}_\mu = \delta^A_\mu$, $e^a{}_\mu = \delta^a_\mu$, and $c_{\mu\nu\rho} = 0$, $\lambda_{abcd} = 0$.
The Killing spinor equations of the maximally supersymmetry version of M2NC in this case are:
\be
\partial_A \epsilon_- + \gamma_A \eta_{-} = 0 \,,\quad \partial_a \epsilon_- = 0 \,,\quad
\partial_A \epsilon_+ + \rho_{+A} = 0 \,,\quad
\partial_a \epsilon_+ = \tfrac12 \gamma_a \eta_- \,.
\label{FLATSPACEKS}
\ee
If $\eta_- = \rho_{+A} = 0$, the only solution is $\epsilon_\pm$ constant, which would give 32 independent constant Killing spinors as for 11-dimensional Minkowski spacetime in the usual 11-dimensional SUGRA.
However, with these parameters present more solutions can be found.
By taking $\partial_a$ derivatives of the first and fourth equations of \eqref{FLATSPACEKS}, and using $\partial_a \epsilon_- = 0$, it follows that $\partial_a \eta_- = 0$ and $\partial_a \partial_b \epsilon_+ = 0$.
The general solution for $\epsilon_+$ and $\rho_{+A}$ then takes the form:
\be
\epsilon_+ = \theta_+ (x^A) + \tfrac12 x^a \gamma_a \eta_-(x^A) \,,\quad
\rho_{+A} = - \partial_A \epsilon_+ \,,
\ee
subject to the condition $\gamma^A \rho_{+A} = 0$ which implies 
\be
\gamma^A \partial_A \theta_+ = 0 \,,\quad
\gamma^A \partial_A \eta_- = 0 \,.
\ee
Given an $\eta_-$ satisfying this equation, $\epsilon_-$ still has be determined from \eqref{FLATSPACEKS}.
As noted in \cite{Bergshoeff:2024nin}, there is one obvious solution with $\eta_- = 0$, namely 
$\epsilon_-$ constant and $\epsilon_+ = \theta_+(x^A)$ an arbitrary solution of the massless longitudinal Dirac equation. This gives infinitely many Killing spinors.

It is therefore somewhat misleading to think in relativistic terms with M2NC SUGRA solutions being fully or half- or quarter-BPS, etc., as this terminology does not easily accommodate infinite-dimensional symmetry enhancements.
It would be desirable to have a more stringent physical criteria for enumerating the preserved supercharges of an M2NC solution.
Note that symmetry enhancement in non-Lorentzian theories -- including field theories -- is not uncommon, see for instance various relevant examples in \cite{Batlle:2016iel, Blair:2020gng, Lambert:2024uue, Lambert:2024yjk, Lambert:2024ncn}, and the solutions I will analyse in this paper fit into this trend.

\subsection{A simple solution: longitudinal AdS${}_3$} 
\label{sec:solutions}

As a brief warm-up before considering solutions associated with the M2 and M5 branes, I first write down a particularly simple ansatz, which I will use to show that the M2NC SUGRA admits a supersymmetric AdS${}_3$ solution.
This exercise will hint at some general features which will appear throughout the rest of the paper.
Let the 11-dimensional coordinates be written as $x^\mu = (x^i, x^I)$, with $i=0,1,2$. Then consider the following ansatz:
\be
\tau^A = \tau^A{}_i(x^i) \dd x^i \,,\quad e^a = \delta^a_I \dd x^I \,, \quad
c_{\mu\nu\rho} = 0 \,,\quad \lambda_{abcd} = \lambda_{abcd} ( x^i) \,,
\label{longansatz}
\ee
which is directly inspired by an observation appearing in \cite{Bergshoeff:2025grj} in the context of the M5NC SUGRA.\footnote{This ansatz \eqref{longansatz} is also suggestive of an extension where the transverse vielbein depends on $x^I$ giving an M2NC version of a Freund-Rubin compactification\cite{Freund:1980xh}. This may be interesting to explore. The near-horizon versions of the M2 and M5 brane solutions described later naively do not fit into such an ansatz as they depend solely on longitudinal or transverse coordinates.}
The ansatz \eqref{longansatz} embeds a three-dimensional geometry into the longitudinal sector of the M2NC geometry, and $\tau^A{}_i$ can be treated as a standard three-dimensional vielbein.
The ansatz \eqref{longansatz} obeys all constraints of the maximally supersymmetric version of the M2NC SUGRA, and hence satisfies all equations of motion identically apart from the Poisson equation \eqref{Poisson}.
This equation becomes
\be
 \Rl_{AB}{}^{AB} + \tfrac{1}{4!} \lambda_{abcd} \lambda^{abcd} = 0 \,,
\label{3P}
\ee
in which $\Rl_{AB}{}^{AB}$ is the Ricci scalar for the three-dimensional geometry with metric $\tau_{ij} = \tau^A{}_i \tau^B{}_j \eta_{AB}$.
Given a geometry whose Ricci scalar is never positive, $R_{AB}{}^{AB}(x^i) \leq 0$, this equation can always be satisfied by taking (for example) $\lambda_{3456} = - \lambda_{78910} = \pm \tfrac{1}{\sqrt{2}} \sqrt{-\Rl_{AB}{}^{AB}}$. 

A special case is constant $\lambda_{abcd} \neq 0$, which singles out locally AdS${}_3$ geometries, with constant negative curvature. 
Then AdS${}_3$ is a solution, and so is the BTZ black hole \cite{Banados:1992wn}, which can be regarded as a particular quotient of global AdS${}_3$.
It is perhaps surprising to find a black hole solution in a non-relativistic theory of gravity.
Practically, this is allowed here because the three-dimensional longitudinal geometry retains an $\mathrm{SO}(1,2)$ Lorentzian structure.
In \cite{Bergshoeff:2025grj}, the six-dimensional Schwarzschild black hole is noted to appear as a similarly longitudinal solution of the M5NC SUGRA.
Perhaps these are not the non-relativistic black holes that one may be looking for, but it would be interesting to understand their role in these limits more thoroughly. Doing so is beyond the scope of the present paper.

Let me instead consider the plain AdS${}_3$ case, and choose coordinates $x^i = (x^\alpha, r)$ such that
\be
\tau^\alpha = \left(\tfrac{r}{L}\right) \dd x^\alpha \,,\quad \tau^2 = \left(\tfrac{L}{r}\right) \dd r \,,\quad
\lambda_{3456} = - \lambda_{78910} = \pm \sqrt{3} / L \,,
\label{adschoice}
\ee
where $L$ is the AdS radius, $\Rl_{AB}{}^{AB} = - 6 / L^2$, and \eqref{3P} is solved.
Can this be a supersymmetric solution? 
The Killing spinor equations follow from setting the supersymmetry transformations \eqref{maxsusytransfs} to zero.
This gives the set of equations
\begin{subequations} 
\label{KSADS}
\begin{align}
0 & = D_i \epsilon_- + \tau^A{}_i \gamma_A \eta_-  \,,\quad
0  = \partial_I \epsilon_-  \,,
\label{KSADSminus}\\[4pt]
0 & = D_i \epsilon_+ +  \tau^A{}_i \rho_{+A}  \mp \tfrac{\sqrt{3}}{6L} \tau^A{}_i \gamma_A \gamma^{3456} \epsilon_+ \,,\quad
0 = \partial_I \epsilon_+ - \tfrac12 \gamma_I \eta_- \mp \tfrac{\sqrt{3}}{4L} \gamma^{3456} \gamma_I \epsilon_- \,,
\label{KSADSplus}
\end{align}
\end{subequations} 
where $D_i$ is the AdS spin connection, which for the vielbein choice of \eqref{adschoice} acts as $D_\alpha \epsilon = \partial_\alpha \epsilon + \tfrac{r}{2L^2} \gamma_\alpha \gamma_r \epsilon$, $D_r \epsilon = \partial_r \epsilon$ (here I relabelled $\gamma_2$ as $\gamma_r$).
In writing down these equations, I used the fact that for a spinor $\chi_+$ such that $\chi_+ = \gamma_{012} \chi_+$, $\gamma^{78910} \chi_+ = - \gamma^{012 789 10} \chi_+ =- \gamma^{3456}\gamma^{012 3456789 10} \chi_+ = - \gamma^{3456} \chi_+$.

The equations \eqref{KSADS} imply firstly that $\epsilon_-$ and hence $\eta_+$ are independent of $x^I$, so that $\epsilon_+$ is given by:
\be
\epsilon_+ = \tilde \epsilon_+(x^i) + x^I \left(
\tfrac12 \gamma_I \eta_-  \pm \tfrac{\sqrt{3}}{4L} \gamma^{3456} \gamma_I \epsilon_- 
\right)  \,.
\label{epspads}
\ee 
Solving for $\rho_{+A}$ one finds
\be
\rho_{+A} = - \tau^i{}_A D_i \epsilon_+ \pm \tfrac{\sqrt{3}}{6L} \gamma_A \gamma^{3456} \epsilon_+ \,,
\ee
and condition that $\rho_{+A}$  is gamma traceless is then:
\be
\gamma^A \rho_{+A} = 0
\Rightarrow  \gamma^A \tau^i{}_A D_i \epsilon_+ \mp \tfrac{\sqrt{3}}{2L}  \gamma^{3456} \epsilon_+ = 0 \,,
\label{AdSDirac}
\ee
which is effectively the Dirac equation in AdS${}_3$ with a mass term.
Then one still has to solve the first equation of \eqref{KSADSminus} (involving $\epsilon_-$ and $\eta_-$) as well as the equations resulting from inserting \eqref{epspads} into \eqref{AdSDirac} (one equation involving $\tilde \epsilon_+$ and one involving $\epsilon_-$, $\eta_-$).

It would be interesting to find the complete set of solutions to these equations.
Here I limit myself to finding infinitely many. 
I take $\epsilon_- = 0 = \eta_-$.
Then, $\partial_I \epsilon_+ = 0$ and I only need to solve the Dirac equation \eqref{AdSDirac} for $\epsilon_+ = \epsilon_+(x^\alpha,r)$.
This equation is: 
\be
\tfrac{L}{r} \gamma^\alpha \partial_\alpha \epsilon_+ 
+ \tfrac{1}{L} \gamma_r \left( r \partial_r \epsilon_+ + ( 1 \mp \tfrac{\sqrt{3}}{2}  \gamma_{01 3456} )\epsilon_+ \right) 
= 0 \,.
\ee
To solve this, assume that $\gamma^\alpha \partial_\alpha \epsilon_+=0$.
The $r$ dependence can then be determined, obtaining: 
\be
\epsilon_+ = r^{ - 1 \pm \tfrac{\sqrt{3}}{2} } \tfrac{1}{2} ( 1 + \gamma_{01 3456} ) \theta_+(x^\alpha)  
+ r^{ - 1 \mp  \tfrac{\sqrt{3}}{2} } \tfrac{1}{2} ( 1 - \gamma_{01 3456} ) \theta_+(x^\alpha)  \,,
\ee
with $\theta_+(x^\alpha)$ an arbitrary solution of the $(1+1)$-dimensional Dirac equation.

It is useful to write down a general solution of this equation in the following manner \cite{Blair:2020gng}.
Let $\theta$ denote an arbitrary (32 component) spinor, which in the M2NC SUGRA will obey some further projection condition. 
Switch to lightcone coordinates, $x^\pm = \tfrac12 (x^0\pm x^1)$, such that the equation $\gamma^\alpha \partial_\alpha \theta = 0$ becomes $(\gamma^+ \partial_+ + \gamma^- \partial_- ) \theta = 0$.
As $(\gamma^+)^2 = (\gamma^-)^2 = 0$, this in fact gives two independent equations $\partial_+ ( \gamma^+ \theta) = 0$, $\partial_- ( \gamma^- \theta) = 0$.
The general solution is therefore
\be
\theta = \theta^{(0)} + \gamma^+ \chi(x^+) + \gamma^- \tilde \chi(x^-)\,,
\label{thetasolnlc}
\ee
with $\theta^{(0)}$ constant and $\chi$, $\tilde \chi$ spinors which are arbitrary non-constant functions of the lightcone coordinates, corresponding to the usual left- and right-moving chiral spinors in $(1+1)$ dimensions.

It follows that this longitudinal AdS${}_3$ solution of M2NC is a supersymmetric background -- with an infinite-dimensional supersymmetry enhancement.

It's worthwhile to comment on an alternative approach to solving these Killing spinor equations.
This would be to choose $\eta_-$ such that the equations satisfied by Killing spinors on AdS (or on products of AdS and other spaces) are recovered.
For example, let $M$ be a matrix such that $(\gamma_r M)^2 = 1$ and $(\gamma_r M) (\gamma_\beta \gamma_r) = - (\gamma_\beta \gamma_r) (\gamma_r M)$.
The simplest choice is $M=1$, but another would be $M=\gamma_r \gamma_{013456}$.
Then for $\eta_ - = - \tfrac{1}{2L} M \epsilon_-$, \eqref{KSADSminus} becomes 
\be
D_i \epsilon_-  = \tfrac{1}{2L} \tau^A{}_i \gamma_A M \epsilon_- \,,
\ee
which can be explicitly solved following \cite{Lu:1998nu} to obtain:
\be
\epsilon_- = r^{1/2} \tfrac12 ( 1 + \gamma_r M) \epsc_- 
+ \left(r^{-1/2} + \tfrac{1}{L^2} r^{1/2} x^\alpha \gamma_\alpha \gamma_r  \right) \tfrac12 ( 1 - \gamma_r M) \epsc_- \,,
\label{ADSeps}
\ee
depending on an arbitrary constant spinor $\epsc_-$.
While this seems promising, one then has to ensure the Dirac equation \eqref{AdSDirac} is solved, with $\epsilon_+$ given by \eqref{epspads} involving the above $\epsilon_-$ and $\eta_-$.
This turns out only to be possible if $\epsilon_- = 0$.
So this longitudinally embedded AdS${}_3$ appears not to give a supersymmetric solution with Killing spinors based on those of AdS.
However later on I will have success with the above technique when considering solutions based on a limit of the M5 brane geometry.

\section{M2 solutions}
\label{sec:M2} 

I now come to the membrane Newton-Cartan limit of the M2 brane supergravity solution, corresponding to the following configuration:

\begin{center}
\begin{tabular}{c|c|c|c|c|c|c|c|c|c|c|c|}
 & $0$ &  $1$ & $2$ & $3$ & $4$ & $5$ & $6$ & $7$ & $8$ & $9$ & ${10}$ \\
M2NC limit & $\times$ & $\times$&  $\times$ & --& --& --& --& --&-- &-- & -- \\
M2 solution & $\times$  &-- &-- &$\times$ & $\times$ &-- &-- &-- &-- & -- &--\\
\end{tabular}
\end{center}
\noindent Accordingly, I write the solution as:
\be
\label{M2}
\begin{split} 
\dd s^2  & = H^{-2/3} ( - (\dd X^0)^2 + \dd X^m \dd X^m ) 
+ H^{1/3} ( \dd X^\alpha \dd X^\alpha + \dd X^I \dd X^I )  \,,\\[4pt]
C_{(3)} & = H^{-1} \dd X^0 \wedge \dd X^3 \wedge \dd X^4 \,,\\[4pt]
H & = 1 + \left( \frac{L}{R} \right)^6 \,,\quad R^2 \equiv X^\alpha X^\alpha + X^I X^I \,,\quad  L^6 \sim N \ell_p^6 \,,
\end{split}
\ee
where $\alpha=1,2$, $m=3,4$ and $I=5,\dots,10$. Here $N$ denotes the number of M2 branes, and $\ell_p$ the Planck length.
To then express this background in the form \eqref{defmnc} from which the M2NC limit can be taken, I perform  the following rescaling 
\be
(X^0, X^\alpha) = c ( x^0, x^\alpha) \,,\quad
(X^m, X^I) = c^{-1/2} (x^m, x^I) \,,
\label{platonicmnc}
\ee
so that \eqref{M2} becomes:
\be
\label{M2platonic}
\begin{split} 
\dd s^2  & = 
c^2 \left(
- H^{-2/3} (\dd x^0)^2 + H^{1/3} \dd x^\alpha \dd x^\alpha
\right) 
+ c^{-1} \left( 
H^{-2/3} \dd x^m \dd x^m  
+ H^{1/3} \dd x^I \dd x^I \right)  \,,\\[4pt]
C_{(3)} & = H^{-1} \dd x^0 \wedge \dd x^3 \wedge \dd x^4 - c^3 \dd x^0 \wedge \dd x^1 \wedge \dd x^2 \,.
\end{split}
\ee
Here I introduced manually the divergent term in $C_{(3)}$.
Crucially this term is closed and so does not modify the field strength, hence the background \eqref{M2platonic} still solves the supergravity equations of motion. 
It is clear here from the terms in the metric proportional to $c^2$ that the piece added to the three-form is indeed $-c^3 \tau^0 \wedge \tau^1 \wedge \tau^2$.
This, together with the finiteness of the remaining part of the three-form, fixes the allowed configuration of the limit, which can be noted to descend from a 1/4 BPS M2-M2 configuration.\footnote{
The only other possibility is a limit where all M2NC longitudinal directions overlap with the M2 worldvolume directions.
This either leads to the near-horizon limit relevant for holography \cite{Blair:2024aqz}, or else
if accompanied by rescaling that sends the number of branes involved to zero, to an apparently non-trivial M2NC geometry (as pointed out in the analogous string example \cite{Avila:2023aey}).
This however is dilatation equivalent to the flat M2NC geometry, commensurate with the fact that no branes are present.}

However, to ensure that the limit of \eqref{M2platonic} is a genuine M2NC geometry, the harmonic function $H$ also has to be finite.
Clearly, \eqref{platonicmnc} produces
\be
\label{Hplatonic}
H = 1 + \frac{L^6}{( c^2 x^\alpha x^\alpha + c^{-1} x^I x^I)^3} \,,
\ee
which as it stands produces $H=1$ in the limit. 
To obtain a non-trivial background, some additional manipulation is needed.
This sets up the choice \cite{Blair:2024aqz, Lambert:2024ncn} between \emph{scaling} or \emph{smearing}, both of which effectively modify the constant $L$ such that a finite and non-trivial $H$ is produced in the limit.

\subsection{Scaled}
\label{sec:M2scaled} 

I first discuss the mechanism where $L$ is scaled.
The M2 solution was treated in this way in \cite{Lambert:2024uue}.
Given that the Planck length is kept fixed in the conventions with which I define the M2NC limit, rescaling $L$ is equivalent to rescaling the number of M2 branes of the background.
This scaled limit extends \eqref{platonicmnc} as follows:
\be
(X^0, X^\alpha) = c ( x^0, x^\alpha) \,,\quad
(X^m, X^I) = c^{-1/2} (x^m, x^I) \,,\quad
N = c^6 n \,.
\ee
This leads to a membrane Newton-Cartan geometry defined by:
\be
\label{M2scaledansatz} 
\begin{split} 
\tau^0 & = H^{-1/3} \dd x^0 \,,\quad
\tau^\alpha  = H^{1/6} \dd x^\alpha \,,\quad
e^m = H^{-1/3} \dd x^m \,\quad
e^I = H^{1/6} \dd x^I \,,\\[4pt]
c_{(3)} & = H^{-1} \dd x^0 \wedge \dd x^3 \wedge \dd x^4 \,,\quad \lambda_{abcd}=0\,,
\end{split}
\ee
with $H= 1 + \frac{\ell^6}{(x^\alpha x^\alpha)^3}$, $\ell^6 \sim n \ell_p^6$.
It's worth mentioning that using the dilatation symmetry, I could express the background \eqref{M2scaledansatz} in a more symmetric form:
\be
\begin{split} 
\tau^0 & = \dd x^0 \,,\quad
\tau^\alpha  = H^{1/2} \dd x^\alpha \,,\quad
e^m = H^{-1/2} \dd x^m \,\quad
e^I = \dd x^I \,,\\[4pt]
c_{(3)} & = H^{-1} \dd x^0 \wedge \dd x^3 \wedge \dd x^4 \,,\quad \lambda_{abcd} = 0 \,.
\end{split}
\label{myansatzdil}
\ee
Nonetheless I will continue to use the original expression \eqref{M2scaledansatz}.
In \cite{Bergshoeff:2024nin}, this background was shown to obey the constraints \eqref{maxcons} needed by the maximally supersymmetric version of the M2NC SUGRA.
In fact, these constraints together with the equations of motions were satisfied for any background of the form \eqref{M2scaledansatz} with $H$ an arbitrary function of the coordinates $x^\alpha$.
It was also shown that functions of the form $H = A (x^\alpha x^\alpha)^B$, $A,B$ constant, gave supersymmetric backgrounds, admitting a Killing spinor solution with 32 supercharges. This of course includes, for $B=-3$, the M2NC limit of the near-horizon limit of the M2 solution.

In \cite{Bergshoeff:2024nin}, it was not shown that the solution obtained was unique, nor were the Killing spinor equations solved for arbitrary $H$.
This motivates revisiting the Killing spinor equations.
I will in fact show that for general $H$, I can find a solution with 24 independent Killing spinors, and that for $\ln H$ harmonic, I can find infinitely many solutions.

I first write down the system of Killing spinor equations for the background \eqref{M2scaledansatz}, taken from \cite{Bergshoeff:2024nin}:
\begin{subequations}
\begin{align}
\label{minusKSa}
0 & = \partial_0 \epsilon_- -\tfrac16 H^{-1/2} \partial_\alpha \ln H \gamma_0 \gamma^\alpha \epsilon_- + H^{-1/3} \gamma_0 \eta_- \,,
\\[4pt]
\label{minusKSb}
0 & = \partial_\alpha \epsilon_- +  \epsilon_{\alpha \beta} \partial^\beta \ln H ( \tfrac{1}{12} \gamma_{12} + \tfrac14 \gamma_{34} ) \epsilon_-  + H^{1/6} \gamma_\alpha \eta_- \,,
\\[4pt] 
\label{minusKSc}
0 & = \partial_m \epsilon_-\,,
\quad
0  = \partial_I \epsilon_-\, \,,\\[4pt]
\label{plusKSa}
0 & = \partial_0 \epsilon_+ -\tfrac16 H^{-1/2} \partial_\alpha \ln H \gamma_0 \gamma^\alpha \epsilon_+ + H^{-1/3} \rho_{+0}\,,
\\[4pt]
\label{plusKSb}
0 & = \partial_\alpha \epsilon_+ +  \epsilon_{\alpha \beta} \partial^\beta \ln H ( \tfrac{1}{12} \gamma_{12} + \tfrac14 \gamma_{34} ) \epsilon_+  + H^{1/6} \rho_{+\alpha}\,,
\\[4pt]
\label{plusKSc}
0 & = \partial_m \epsilon_+ 
+ H^{-1/2} \tfrac12 ( \tfrac13  \partial^\alpha\ln H \gamma_\alpha \gamma_m 
+ \tfrac12 \epsilon^{\alpha \beta} \epsilon_{mn} \partial_\beta \ln H \gamma_\alpha \gamma^n ) \epsilon_- - \tfrac12 H^{-1/3} \gamma_m \eta_- \,,\\[4pt]
0 & = \partial_I \epsilon_+ - \tfrac{1}{12} \partial^\alpha \ln H \gamma_\alpha \gamma_I \epsilon_-
- \tfrac12 H^{1/6} \gamma_I \eta_- \,.
\label{plusKSd}
\end{align}
\end{subequations} 
To begin solving these equations, firstly note that $\eta_-$ can always be determined from \eqref{minusKSa} as:
\be
\eta_- = \tfrac16 H^{-1/6} \partial_\alpha \ln H \gamma^\alpha \epsilon_-
 + H^{1/3} \gamma_0 \partial_0 \epsilon_- \,.
\label{M2eta} 
\ee 
Similarly, \eqref{plusKSa} and \eqref{plusKSb}, can be solved for $\rho_{+A}$.
Requiring $\gamma^A \rho_{+A} = 0$ leads to the following differential equation:
\be
0 = H^{1/2} \gamma^0 \partial_0 \epsilon_+ 
+ \gamma^\alpha \left( \partial_\alpha \epsilon_+ %+ \tfrac16 \partial_\alpha \ln H \epsilon_+ 
+ \partial_\alpha \ln H ( - \tfrac{1}{12} + \tfrac{1}{4} \gamma_{034} )\epsilon_+ \right)  = 0 \,.
\label{M2rho}
\ee
In order to solve the remaining equations, I will make the simplifying assumption that $\partial_0 \epsilon_- = 0$.
This leads to a more tractable system of equations.
In particular, equation \eqref{plusKSd} becomes simply $\partial_I \epsilon_+ = 0$, while \eqref{minusKSb} can be written as
\be
0 = \partial_\alpha \epsilon_- + \tfrac16 \partial_\alpha \ln H \epsilon_- + \tfrac14  \epsilon_{\alpha \beta} \partial^\beta \ln H\gamma_0 (1-\gamma_{034} ) \epsilon_- \,.
\label{minusKSb_eta}
\ee
To solve this equation I rewrite
%\be
$\epsilon_- = H^{-1/6} ( \chi_-(x^\alpha) + \tilde \chi_-(x^\alpha)  )$,
decomposing in terms oppositely projected spinors, $\chi_- = \gamma_{034} \chi_-$, $\tilde \chi_- = - \gamma_{034} \tilde \chi_-$.
Then, inserting into into \eqref{minusKSb_eta}, I find $\partial_\alpha \chi_- = 0$ as well as 
\be
\label{tildechieqn}	
\partial_\alpha  \tilde \chi_- = - \tfrac{1}{2} \epsilon_{\alpha \beta} \partial^\beta \ln H \gamma_0 \tilde \chi_- \,.
\ee
It immediately follows that $\partial^\alpha \ln H \partial_\alpha \tilde \chi_- = 0$; differentiating both sides of \eqref{tildechieqn} with $\epsilon^{\alpha \beta} \partial_\beta$ one then finds necessarily that
\be
\partial^\alpha \partial_\alpha \ln H = 0 \,,
\label{CleverHM2}
\ee
unless $\tilde \chi_- = 0$.
The equation \eqref{CleverHM2} imposes that $\ln H$ is harmonic in the $x^\alpha$ directions: this condition was highlighted in \cite{Bergshoeff:2024nin} as picking out the family of solutions which include the near-horizon form of the scaled M2 solution.
More generally, suppose that $\ln H$ is harmonic, and that we can find a function $\phi(x^\alpha)$ which is a harmonic conjugate of $\ln H$, obeying:
\be
\partial_\alpha \phi  = \epsilon_{\alpha \beta} \partial^\beta \ln H \,.
\label{Hconj}
\ee
Then the solution of \eqref{tildechieqn} is $\tilde \chi_- = \exp \left(-\tfrac12 \phi(x^\alpha) \gamma_0 \right) \tilde \chi_-^{(0)}$ for a constant spinor $\tilde \chi^{(0)}_-$.
The full solution for $\epsilon_-$ is thus: 
\be
\epsilon_- = H^{-1/6} \left( \tfrac12 (1+\gamma_{034} ) + \exp \left(-\tfrac12 \phi(x^\alpha) \gamma_0 \right) \tfrac12 (1-\gamma_{034} ) \right) \epsc_- \,,
\label{epsminusM2soln} 
\ee
with $\epsc_-$ constant. If $H$ does not obey \eqref{CleverHM2}, then the allowed solution restricts to the case $\epsc_- = \gamma_{034} \epsc_-$.
Observe that for the family of backgrounds with $H  = A (x^\alpha x^\alpha)^B$ that 
\be
\phi = - 2 B \theta \,,\quad \theta \equiv \tan^{-1} ( x^2/x^1) \,,
\ee
with $\theta$ being the angular coordinate in polar coordinates.
Using this, it can be checked that \eqref{epsminusM2soln} agrees with the solution for $\epsilon_-$ found in \cite{Bergshoeff:2024nin}.

Next I turn to the equations for $\epsilon_+$.
Firstly, I consider \eqref{plusKSc}. Inserting $\eta_-$ as given in \eqref{M2eta} (with the assumption $\partial_0 \epsilon_- = 0$), this equation is solved simply by
\be
\epsilon_+ = \tilde \epsilon_+(x^0,x^\alpha) - \tfrac14 H^{-1/2} \partial^\alpha \ln H \gamma_\alpha x^m \gamma_m ( 1 - \gamma_{034} ) \epsilon_- \,.
\ee
A short calculation shows that the $x^m$-dependent terms in $\epsilon_+$ cancel in the equation \eqref{M2rho} that follows from the condition $\gamma^A \rho_{+A} = 0$, assuming that $\ln H$ is harmonic.
It remains to find solutions where $\epsilon_+$ depends only on $x^0$ and $x^\alpha$.
I will again make the simplifying assumption that $\partial_0 \epsilon_+ = 0$, and insert the decomposition
\be
\epsilon_+ (x^0,x^\alpha) = 
\chi_+(x^\alpha) + \tilde \chi_+(x^\alpha)
 \,,\quad
\chi_+ = \gamma_{034} \chi_+ \,,\quad
\tilde \chi_+ = - \gamma_{034} \tilde \chi_+ \,,
\ee
into \eqref{M2rho}.
I then find the equations:
\be
\gamma^\alpha ( \partial_\alpha \chi_+ + \tfrac16 \partial_\alpha \ln H \chi_+ ) = 0 
\,,\quad
\gamma^\alpha ( \partial_\alpha \tilde \chi_+ - \tfrac13 \partial_\alpha \ln H \tilde \chi_+ ) = 0 \,.
\label{chipluseqns}
\ee
My trick to solve these is to convert them into projections that annihilate $\chi_+$ and $\tilde \chi_+$.
This goes as follows.
Assuming again that $\ln H$ is harmonic and that the conjugate $\phi$ can be found, I make the following ansatz:
\be
\chi_+ = H^\kappa \exp ( \lambda \phi \gamma_0 ) \chi_+^{(0)}  \,,\quad
\tilde \chi_+ = H^{\tilde \kappa} \exp(\tilde \lambda \phi \gamma_0 ) \tilde \chi_+^{(0)} \,,
\ee
in terms of some numbers $\kappa,\tilde\kappa$, $\lambda,\tilde\lambda$ and constant spinors $\chi_+^{(0)}$, $\tilde \chi_+^{(0)}$.
Inserting this ansatz into \eqref{chipluseqns} one finds
\be
\partial_\alpha \ln H ( ( \kappa+ \tfrac16 ) \gamma^\alpha- \lambda \epsilon^{\alpha \beta} \gamma_\beta \gamma_0 ) \chi_+ = 0 
= \partial_\alpha \ln H ( ( \tilde \kappa- \tfrac13 ) \gamma^\alpha- \tilde \lambda \epsilon^{\alpha \beta} \gamma_\beta \gamma_0 ) \tilde \chi_+\,.
\ee 
Using $\gamma^\alpha \gamma_{012}  = \epsilon^{\alpha \beta} \gamma_\beta \gamma_0$, and the fact that $\chi_+$ and $\tilde \chi_+$ are by definition eigenspinors of $\gamma_{012}$ with eigenvalue $+1$, these equations are satisfied for $\lambda = \kappa + \tfrac16$ and $\tilde \lambda = \tilde \kappa-\tfrac13$.
There is therefore an infinite family of solutions for $\epsilon_+$ of the form:
\be
\begin{split} 
\epsilon_+ & =
H^{\lambda - \tfrac16}\exp ( \lambda \phi  \gamma_0) \tfrac12 (1 + \gamma_{034}) \epsc_+ 
+ 
H^{\tilde \lambda + \tfrac13} \exp ( \tilde \lambda  \phi\gamma_0) \tfrac12 (1 - \gamma_{034}) \epsc_+
\\[4pt] & \qquad - \tfrac14 H^{-1/2} \partial^\alpha \ln H \gamma_\alpha x^m \gamma_m ( 1 - \gamma_{034} ) \epsilon_- \,,
\end{split} 
\label{epsplusM2soln}
\ee
for arbitrary numbers $\lambda, \tilde \lambda$, with $\epsc_+$ constant, $\epsilon_-$ given by \eqref{epsminusM2soln}, $H$ obeying \eqref{CleverHM2} and $\phi$ obeying \eqref{Hconj}.
The solution found in \cite{Bergshoeff:2024nin} has $\lambda = \tfrac13$ and $\tilde \lambda = - \tfrac16$.\footnote{The fact these general solutions were overlooked in \cite{Bergshoeff:2024nin} can perhaps be attributed to the setting in of a satisfied relativistic bias once the magic number 32 of Killing spinor solutions had been found.}

If $H$ does \emph{not} obey \eqref{CleverHM2}, which in particular includes the scaled M2 solution in its original non-near-horizon form, then only the case with $\lambda = 0 = \tilde \lambda$ is still a solution.
This means the full solution obtained by the above method for general $H$ not obeying \eqref{CleverHM2} has the form: 
\be
\epsilon_+  = H^{-1/6} \tfrac12 ( 1+ \gamma_{034} ) \epsc_+ 
+ H^{1/3} \tfrac12 ( 1 - \gamma_{034} ) \epsc_+ \,,\quad 
\epsilon_-  = H^{-1/6}  \tfrac12 (1+\gamma_{034} ) \epsc_- \,,
\ee
with $\eta_-$ as in \eqref{M2eta}, $\rho_{+A}$ determined from \eqref{plusKSa} and \eqref{plusKSb}, and 
$\epsc_\pm$ constant.
Curiously, this means that the M2NC limit of the original M2 solution admits (at least) 24 Killing spinors in the maximally supersymmetric version of the M2NC SUGRA.
The Killing spinors with power $H^{-1/6}$ are inherited from those of the usual M2 solution.
The `additional' ones with power $H^{1/3}$ are allowed here owing to the presence of the additional fermionic shift symmetries.
Taking the near-horizon limit of the M2 solution, one finds a supersymmetry enhancement to the solution with $\epsilon_\pm$ given by \eqref{epsminusM2soln} and \eqref{epsplusM2soln}.

It is not impossible that there are further solutions to the Killing spinor equations with $\partial_0 \epsilon_- \neq 0$.
The equations in this case are not obviously inconsistent, but neither do they obviously lead to solutions. 
It can be checked, for instance, that $\epsc_\pm$ cannot be enhanced to functions of $x^0$.
Note that the enhancements appearing in the flat space background discussed in section \ref{sec:fermions} and the AdS background of section \ref{sec:solutions} (as well as the enhancements which will be seen in section \ref{sec:M5} for the M5 brane) rely on the possibility of imposing a Lorentzian signature Dirac equation in the longitudinal directions. This is not possible here, as the longitudinal time and spatial coordinates are treated separately.

The above derivations show that any background of the form \eqref{M2scaledansatz} with $\ln H$ harmonic is a solution of the maximally supersymmetric version of M2NC SUGRA admitting an infinite number of Killing spinors (assuming that there exists a $\phi$ satisfying \eqref{Hconj}). While this is intriguing, it may seem like an embarrassment of riches.
It is therefore interesting to contrast it with the M2 solution obtained by smearing, which I study next.

\subsection{Smeared}
\label{sec:M2smeared} 

Now I discuss the alternative limit of the M2 brane where the solution is smeared on the longitudinal directions of the limit.
The solution smeared on the $X^\alpha$ directions has the same form as \eqref{M2} except now the harmonic function becomes:
\be
\begin{split} 
H  = 1 + \left( \frac{\tilde L}{R} \right)^4 \,,\quad R^2 \equiv X^I X^I \,,\quad  \tilde L^4 \sim \frac{1}{V} N \ell_p^6 \,.
\end{split}
\ee
where $V$ denotes the volume of the $X^\alpha$ directions.
The rescaling \eqref{platonicmnc} then implies $V = c^2 v$.
This leads to an M2NC geometry where the fields have the same form as in \eqref{M2scaledansatz} but now the harmonic function is $H = 1 + \ell^4/(x^Ix^I)^2$
with $\ell^4 \sim \frac{1}{v} N \ell_p^6$.
This is localised in the transverse directions.

With this choice of harmonic function, it can easily be checked that while $f_{abcd}$ and $T_{ab}{}^A$ vanish, $f_{Aabc}$ and $T_{a\{AB\}}$ are non-zero.
Referring to the defining constraints \eqref{maxcons} and \eqref{halfcons}, this can only be a solution in the half-maximally supersymmetry version of M2NC SUGRA! 
I now verify that this is indeed a solution, and that it admits 8 independent Killing spinors.

In fact, in a spirit of experimentation, I will examine a more general family of backgrounds, which `combine' the results of the scaled and smeared limits by taking as an ansatz the M2NC background \eqref{M2scaledansatz} with the assumption that 
\be
H  = H(x^\alpha, x^I) \,,
\label{myansatzH}
\ee
allowing for the fields to depend on all the coordinates $x^\alpha$ and $x^I$, which are the coordinates originally transverse to the relativistic M2 brane solution.
It is straightforward to build the torsions and anholonomy coefficients of \eqref{defTOmega}:
\begin{subequations} \label{buildsmearedM2}
\begin{align}
T_{BC}{}^A :\quad & T_{\alpha 0}{}^0 = - \tfrac13 H^{-7/6} \partial_{\alpha} H \,,\quad
T_{\beta \gamma}{}^\alpha  = \tfrac{1}{3} H^{-7/6} \partial_{[\beta} H \delta^\alpha_{\gamma]} \,,
\\[4pt]
T_{aA}{}^B: \quad
 & T_{I0}{}^0 = -  \tfrac13 H^{-7/6} \partial_I H\,,\quad
 T_{I\beta}{}^\alpha = \tfrac16 H^{-7/6} \partial_I H \delta_\beta^\alpha \,,
\\[4pt]
\Omega_{Ab}{}^a :\quad & \Omega_{\alpha n}{}^m  = - \tfrac13 H^{-7/6} \delta^m_n \partial_{\alpha} H \,,\quad
 \Omega_{\alpha J}{}^I = \tfrac16 H^{-7/6} \delta^I_J \partial_{\alpha} H \,,
\\[4pt]
\Omega_{bc}{}^a: \quad
 & \Omega_{I n}{}^m = -  \tfrac13 H^{-7/6} \delta_n^m \partial_I H\,,\quad
 \Omega_{JK}{}^I = \tfrac13 H^{-7/6} \partial_{[J} H \delta_{K]}^I \,,
\end{align}
as well as the flat index components of the four-form field strength:
\begin{align}
f_{ABab}: \quad & f_{0\alpha mn} =  H^{-7/6} \partial_\alpha H \epsilon_{mn} \,,\quad
f_{Aabc}: \quad  f_{0 I mn}  = H^{-7/6} \partial_I H \epsilon_{mn} \,.
\end{align} 
\end{subequations} 
Note that $T_{aA}{}^{A} = 0$. 
As expected from the transverse dependence of $H$, $f_{Aabc}$ and $T_a{}^{\{AB\}}$ are non-zero, and this can only be a solution of the half-maximal version of M2NC SUGRA.

The non-zero components of the connections \eqref{connections} are:
\begin{subequations} 
\begin{align}
\omega_\mu{}^{AB} :\quad &
\omega_{\hat 0}{}^{0 \alpha} = - \tfrac13 H^{-3/2} \partial^\alpha H \,,\quad
\omega_{\hat \alpha}{}^{\beta \gamma}= \tfrac13 H^{-1}  \delta^{[\beta}_{\alpha} \partial^{\gamma]} H \,,
\\[4pt] 
\omega_\mu{}^{ab}: \quad &
\omega_{\hat \alpha}{}^{34} = \tfrac12 H^{-1} \epsilon_{\alpha \beta} \partial^\beta H  \,,
\\[4pt]
 & \omega_{\hat m}{}^{nI} = -\tfrac13 H^{-3/2} \delta_m^n \partial^I H\,,\quad
 \omega_{\hat I}{}^{JK} = \tfrac13 H^{-1} \delta_I^{[J} \partial^{K]} H \,,
\\[4pt]
\omega_\mu{}^{Aa}: \quad &
\omega_{\hat m}{}^{\alpha n} =
H^{-3/2} ( 
\tfrac13 \delta^n_m \partial^\alpha H 
+\tfrac12 \epsilon^{\alpha \beta} \epsilon_m{}^n \partial_\beta H 
)
\,,\quad
\omega_{\hat I}{}^{\alpha J} = - \tfrac16 H^{-1} \delta_I^J \partial^\alpha H \,,
\end{align}
with $b_\mu = 0$. Hats are used here on the left-hand side of equations to distinguish the curved index of $\omega$ from the flat ones.
\end{subequations} 
Working systematically through the equations of motion \eqref{EOMCOV}, one can 
verify that these only impose that $H$ is harmonic in the coordinates $x^I$:
\be
\partial^I \partial_I H = 0 \,.
\label{Hharmonic}
\ee
This means in particular that any derivatives of $H$ with respect to the coordinates $x^\alpha$ simply drop out of the equations of motion, leaving the dependence on these coordinates largely unconstrained.
Recall these coordinates are longitudinal to the M2NC limit but transverse to the original M2 appearing in the supergravity solution.
Note though that it is \emph{not} possible to take $H$ to have the functional dependence of the original harmonic function, that is $H \sim 1/(x^\alpha x^\alpha + x^I x^I)^3$, as this does not satisfy \eqref{Hharmonic}. An example of an allowed solution would be $H = A(x^\alpha) + B(x^\alpha) / (x^I x^I)^2$, with $A$ and $B$ some functions of the spatial longitudinal coordinates.

I now examine the Killing spinor equations, which follow by setting to zero the fermionic variations \eqref{halfsusytransfs} of the half-maximal version of the M2NC SUGRA.
Taking longitudinal and transverse projections, and taking into account that $b_\mu=\lambda_{abcd} = 0$ for this background, these can be written as: 
\begin{subequations}\label{KSansatz}
\begin{align} 
\label{KS1}
0& =  \left(
\tfrac12  T_{a}{}^{\{AB\}} \gamma_B \gamma^a 
+\tfrac14 ( \eta^{AB} - \tfrac13 \gamma^A \gamma^B ) \tfrac16 f_{Bcde} \gamma^{cde} \right) \epsilon_+
\,,\\[4pt] 
0 & = \left(e^\mu{}_a \partial_\mu  + \tfrac{1}{4} \omega_a{}^{bc} \gamma_{bc}  
+\tfrac{1}{12} \tfrac{1}{6}  f_{A bcd} ( \gamma_a{}^{bcd} - 6 \delta_a^b \gamma^{cd} ) \gamma^A \right) \epsilon_+  \,,
\label{KS2}
\\[4pt]
0 & = (\tau^{\mu}{}_A \partial_\mu   + \tfrac{1}{4} \omega_A{}^{ab} \gamma_{ab} 
+ \tfrac{1}{4}  \omega_A{}^{BC} \gamma_{BC} )\epsilon_+
+ \rho_{+A}  \,.
\label{KS3}
\end{align}
Here I have also made use of the fact that $\eta_- = 0$ is necessarily implied by the structure of the equation $\delta \psi_{-\mu} = 0$.
\end{subequations}

The first of these equations \eqref{KS1} immediately looks as if it will lead to a projection condition on the Killing spinor.
This is indeed the case: inserting the expressions for $T_a{}^{\{AB\}}$ and $f_{Aabc}$ following from \eqref{buildsmearedM2}, it boils down to the worldvolume projection condition that was already encountered above:
\be
\epsilon_+ = \gamma_{034} \epsilon_+ \,.
\label{projection_condition}
\ee
Using \eqref{projection_condition} in the equation \eqref{KS2} then leads to two simple differential equations: 
\be
\partial_m \epsilon_+ = 0 \,,\quad  \partial_I \epsilon_+ + \tfrac16  \partial_I \ln H \epsilon_+ = 0 \,,
\label{M2smearedHcondition}
\ee
solved by
\be
\epsilon_+(x^\alpha, x^I) = H^{-1/6} ( x^\alpha,x^I)\, \tilde \epsilon_+(x^0,x^\alpha) \,,\quad
\tilde \epsilon_+ = \gamma_{034} \tilde \epsilon_+ \,.
\ee
Inserting this into \eqref{KS3}, and using $\gamma_{12} \tilde \epsilon_+ = \gamma_{34} \tilde \epsilon_+ = - \gamma_0 \tilde \epsilon_+$ results in:
\begin{subequations} \label{longM2KS}
\begin{align}
0& = H^{1/6} \partial_0 \tilde \epsilon_+ - \tfrac16 H^{-1/3} \partial^\alpha H \gamma_{0 \alpha} \tilde \epsilon_+  + \rho_{+0}\,,
\\[4pt]
0&= H^{-1/3} \partial_\alpha \tilde \epsilon_+ - \tfrac16 H^{-1/3} \partial_\alpha \ln H \tilde \epsilon_+  - \tfrac{1}{3} H^{-1/3} \epsilon_{\alpha \beta} \partial^\beta \ln H \gamma_{0}\tilde \epsilon_+
+ \rho_{+\alpha}\,.
\end{align}
\end{subequations} 
In effect, these solve for $\rho_{+A}$, which must obey $\gamma^A \rho_{+A} = 0$.
A short calculation shows that
\be
\gamma^A \rho_{+A} = - H^{-1/6} \gamma^0 \partial_0 \tilde \epsilon_+ 
 - H^{-1/3} \gamma^\alpha \partial_\alpha \tilde \epsilon_+ \,,
 \label{gammatraceM2}
\ee
given that $\tilde \epsilon_+$ is projected. 
As $H$ depends on $x^I$ but $\tilde \epsilon_+$ does not, for \eqref{gammatraceM2} to be zero both terms in the sum have to be zero separately, that is $\partial_0 \tilde \epsilon_+ = 0$ and $\gamma^\alpha \partial_\alpha \tilde \epsilon_+ = 0$.
The only solution is to take $\tilde \epsilon_+ = \epsc_+$ constant.
Thus the full solution obtained is: 
\be
\epsilon_+(x^\alpha, x^I) = H^{-1/6} ( x^\alpha,x^I) \, \epsc_+ \,,\quad
 \epsc_+ = \gamma_{034} \epsc_+ \,,
\label{epsM2}
\ee
with $\rho_{+A}$ determined via \eqref{longM2KS}.
For $H$ independent of $x^\alpha$, the result \eqref{epsM2} is simply the usual Killing spinor solution for the (smeared) M2 brane, here supplemented with the single projection $\epsilon \rightarrow \epsilon_+$ induced by the underlying M2NC limit. 
This is in line with the interpretation of the M2NC limit itself as representing a half-supersymmetric M2 brane oriented in the 012 directions.
This is orthogonal to the physical M2 brane in the 034 directions, leading to an underlying 1/4 BPS configuration.

It can be checked that the Killing spinor equation \eqref{KS1} always implies the projection condition \eqref{projection_condition} as long as $\partial_I H \partial^I H \neq 0$, which is guaranteed unless $H$ depends only on the longitudinal directions $x^\alpha$.
As a result, it appears there is no enhancement of supersymmetry in the near-horizon limit, that is when taking $H = \ell^4 / ( x^I x^I)^2$.
This could also be understood as being inherited from the original relativistic configuration.
The M2 solution smeared on two directions has as its near-horizon limit a geometry conformal to \adsS{4}{5}$\times$ T${}^2$, instead of the maximally supersymmetric \adsS{4}{7}.
Furthermore, the above derivation appears to rule out any non-Lorentzian enhancements. This is essentially because the projection condition \eqref{projection_condition} enforces \eqref{M2smearedHcondition}, thereby fixing both the projection and the $H$ dependence of the Killing spinor.

\section{M5 solutions} 
\label{sec:M5}

Having analysed the M2 brane, it is only natural to consider its electromagnetic dual counterpart, the M5.
The only possible limit that can be taken corresponds to the following configuration:

\begin{center}
\begin{tabular}{c|c|c|c|c|c|c|c|c|c|c|c|}
 & $0$ &  $1$ & $2$ & $3$ & $4$ & $5$ & $6$ & $7$ & $8$ & $9$ & ${10}$ \\
M2NC limit & $\times$ & $\times$&  $\times$ & --& --& --& --& --&-- &-- & -- \\
M5 solution & $\times$  & $\times$ &-- &$\times$ & $\times$ &$\times$ &$\times$  &-- &-- & -- &--\\[4pt]
\end{tabular}
\end{center}

\noindent Accordingly, I write the M5 solution as:
\be
\label{M5}
\begin{split} 
\dd s^2  & = H^{-1/3} (\eta_{\alpha \beta} \dd X^\alpha \dd X^\beta + \dd X^m \dd X^m ) 
+ H^{2/3} ((\dd X^2)^2  + \dd X^I \dd X^I )  \,,\\[4pt]
C_{(6)} & = H^{-1} \dd X^0 \wedge \dd X^1 \wedge \dd X^3 \wedge \dots \wedge \dd X^6 \,,\\[4pt]
H & = 1 + \left( \frac{L}{R} \right)^3 \,,\quad R^2 \equiv X^2 X^2 + X^I X^I \,,\quad  L^3 \sim N \ell_p^3 \,,
\end{split}
\ee
where now $\alpha=0,1$, $m=3,4,5,6$ and $I=7,\dots,10$.
The field strength of the six-form is here given simply by $F_{(7)} = \dd C_{(6)}$ and is related to the electric field strength by $F_{(4)} = - \star F_{(7)}$. 
The four-form is then given explicitly by:
\be
F_{(4)} = \dd X^7 \wedge \dots \wedge \dd X^{10} \frac{\partial H}{\partial X^2} 
- \tfrac16 \epsilon_{IJKL} \dd X^I \wedge \dd X^J \wedge \dd X^K \wedge \dd X^2\frac{\partial H}{\partial X^L} \,.
\label{fourform}
\ee
The M2NC prescription in fact requires \cite{Blair:2021waq} that the six-form potential takes the form:
\be
C_{(6)} = - c^3 c_{(3)} \wedge \tau^0 \wedge \tau^1 \wedge \tau^2 + c_{(6)} \,.
\label{sixformmnc}
\ee
However, it is more straightforward to mostly work with the electric variables below, using \eqref{fourform}. 

\subsection{Scaled} 
\label{sec:M5scaled}

I first consider the limit incorporating the large $N$ scaling, which here corresponds to the prescription:
\be
(X^\alpha, X^2) = c (x^\alpha, u) \,,\quad
(X^m, X^I) = c^{-1/2} (x^m, x^I) \,,\quad
N = c^3 n \,,
\ee
where to marginally improve index aesthetics I relabel the longitudinal coordinate $x^2$ as $u$.
Under this scaling the harmonic function becomes
\be
H = 1 + \frac{\ell^3}{(u^2 + c^{-3} x^I x^I)^{3/2}} \,,
\label{HM5scaled}
\ee
and the electric field strength is: 
\be
F_{(4)} = c^{-3} \dd x^7 \wedge \dots \wedge \dd x^{10} \frac{\partial H}{\partial u} 
- \tfrac16 \epsilon_{IJKL} \dd x^I \wedge \dd x^K \wedge \dd x^L \wedge \dd u\frac{\partial H}{\partial x^L} \,.
\ee
Both terms in this expression are of order $c^{-3}$, as follows by working out the derivatives of \eqref{HM5scaled}.
This means that there is no three-form in the M2NC geometry that arises after taking the limit.
However, there will be a non-trivial Lagrange multiplier field $\lambda_{abcd}$.
As explained in section \ref{sec:bosons}, before taking the limit, this is an auxiliary field determined by its algebraic equation of motion to be $\lambda_{abcd} = c^{3} f^{(-)}_{abcd}$, where $f^{(-)}_{abcd}$ refers to the anti-self-dual projection of the transverse projection of the finite part of the field strength.
Taking this into account leads to an M2NC geometry of the following form:
\be
\begin{split} 
\tau^\alpha & = H^{-1/6} \dd x^\alpha \,,\quad \tau^u= H^{1/3} \dd u \,,\quad
e^m = H^{-1/6} \dd x^m \,,\quad e^I = H^{1/3} \dd x^I \,,\\[4pt]
c_{(3)} & = 0 \,,\quad 
\lambda_{3456}=  - \lambda_{789 10} = - \tfrac12 H^{-4/3} \partial_u H \,,\\[4pt]
H & = 1 + \frac{l^3}{u^3}\,,\quad l^3 \sim n \ell_p^3 \,.
\end{split} 
\label{M5ScaledM2NC}
\ee
As $c_{(3)}$ is zero, the required divergent term in the six-form of \eqref{sixformmnc} is also zero and therefore automatically (and trivially) accounted for.\footnote{In the M2NC SUGRA, it is consistent to have $c_{(6)}$ non-zero with $c_{(3)}$ zero, as the field strength $f_{(7)}$ of the former is not just related to $f_{(4)}$ but also to $\lambda_{abcd}$. This follows by defining $f_{(7)}$ such that $\dd f_{(7)}$ reproduces the left-hand side of the equation of motion \eqref{EomChere}.}

Now I examine the geometry and equations of motion.
The non-trivial components of the torsions and anholonomy coefficients of \eqref{defTOmega} are:
\be
T_{\hat u \hat \alpha}{}^\beta = - \tfrac{1}{6} H^{-7/6} \partial_u H \delta_\alpha^\beta \,,\quad
\Omega_{\hat u \hat n}{}^m = - \tfrac{1}{6} H^{-7/6} \partial_u H \delta_n^m 
\,,\quad 
\Omega_{\hat u \hat J}{}^I = \tfrac{1}{3} H^{-2/3} \partial_u H \delta_J^I\,,
\ee
where hats on the left-hand side of equations here distinguish curved indices of $T$ and $\Omega$.
Therefore one has $T_{ab}{}^A = 0 = T_{aA}{}^B = 0$.
I go on to evaluate the connections \eqref{connections}, obtaining:
\begin{subequations}
\begin{align}
 \omega_\mu{}^{AB} :\quad &  \omega_{\hat \alpha}{}^{\beta u} = - \tfrac{1}{6} H^{-3/2} \partial_u H \delta_\alpha{}^\beta \,,\\[4pt]
   \omega_\mu{}^{Aa} :\quad &  \omega_{\hat m}{}^{u n} = \tfrac16 H^{-3/2} \partial_u H \delta_m^n \,,\quad  \omega_{\hat I}{}^{u J} = -\tfrac13 H^{-1} \partial_u H \delta_I^J \,,
\end{align}
as well as $\omega_\mu{}^{ab} =  0$, $b_{\mu} = 0$.
\end{subequations}
For the curvatures \eqref{improvedR} and \eqref{Poisson} appearing in the constraints and Poisson equation, it then follows that $\breve R_{\mu\nu}{}^{ab} = 0$
while
\be
\breve{R}_{A(a}{}^B{}_{b)} = 
H^{-8/3}
\begin{pmatrix}
(- \tfrac{7}{16} ( \partial_u H)^2 + \tfrac14 H\partial_u^2 H ) \delta_{mn} & 0 \\
0 & (\tfrac{5}{16} ( \partial_u H)^2 - \tfrac14 H\partial_u^2 H ) \delta_{IJ}
\end{pmatrix} \,.
\ee
Then, for $H$ an arbitrary function of the coordinate $u$, the constraints \eqref{maxcons} are satisfied: the constraints \eqref{maxcons_torsion} follow immediately, while the covariant derivative constraints of \eqref{maxcons_curv} reduce simply to the conditions that $\lambda_{abcd}$ and $\breve{R}_{Aa}{}^{Ab}$ are independent of the transverse coordinates, which is obviously true.
Finally, the Poisson equation, $\breve{R}_{Aa}{}^{Aa} + \tfrac{1}{4!} \lambda_{abcd} \lambda^{abcd} = 0$, is seen to hold.
Hence this is a valid solution of the maximally supersymmetric version of M2NC SUGRA.

I now show that it further gives a supersymmetric solution.
\begin{subequations}
\label{KSM5Scaled}
The Killing spinor equations for the background \eqref{M5ScaledM2NC}, obtained by setting the fermionic variations \eqref{maxsusytransfs} to zero,
are:
 \begin{align}
0 & = \partial_m \epsilon_- = \partial_I \epsilon_- \,,\\[4pt]
\label{KSepsminusalpha}
0 & = 
\partial_\alpha  \epsilon_-- \tfrac{1}{12} H^{-3/2} \partial_u H \gamma_\alpha \gamma_u \epsilon_- + H^{-1/6} \gamma_\alpha \eta_-\,,\\[4pt]
\label{KSepsminus2}
0 & = 
\partial_u \epsilon_- + H^{1/3} \gamma_u \eta_-\,,\\[4pt]
0 & = \partial_\alpha \epsilon_+ - \tfrac{1}{12} H^{-3/2} \partial_u H \gamma_\alpha \gamma_u ( 1 - \gamma_{013456} ) \epsilon_+ + H^{-1/6} \rho_{+ \alpha}\,,
\label{KSepsplusalpha}\\[4pt]
0 & = \partial_u \epsilon_+ + \tfrac{1}{12} H^{-1} \partial_u H \gamma_{013456} \epsilon_+ + H^{1/3} \rho_{+u} \,,\label{KSepsplus2}\\[4pt]
0 & = \partial_m \epsilon_+ + H^{-3/2} \partial_u H \gamma_u \gamma_m ( \tfrac{1}{12} - \tfrac{1}{8} \gamma_{013456} ) \epsilon_- 
-\tfrac12 H^{-1/6} \gamma_m \eta_-\,, \label{KSepsplusm}\\[4pt]
0 & = \partial_I \epsilon_+ + H^{-1} \partial_u H \gamma_u \gamma_I (- \tfrac{1}{6} + \tfrac{1}{8} \gamma_{013456} ) \epsilon_- 
-\tfrac12 H^{1/3} \gamma_I \eta_- \,.\label{KSepsplusI}
\end{align} 
\end{subequations}
The parameter $\eta_-$ can be determined from \eqref{KSepsminus2} as
\be
\eta_- = - H^{-1/3} \gamma_u \partial_u \epsilon_- \,,
\label{DetermineEta}
\ee
which substituting into \eqref{KSepsminusalpha} implies
\be
 \partial_\alpha \epsilon_-  
 =  H^{-1/2} \gamma_\alpha \gamma_u  \left( \partial_u \epsilon_- 
 + \tfrac{1}{12} H^{-1} \partial_u H \epsilon_- \right) \,.
 \label{DetermineEpsM}
\ee
Then, given that $\partial_m \epsilon_- = \partial_I \epsilon_- = 0$, the equations \eqref{KSepsplusm} and \eqref{KSepsplusI} can also be solved to determine the linear transverse coordinate dependence of $\epsilon_+$:
\be
\begin{split}
\epsilon_+& = \tilde \epsilon_+ ( x^\alpha, x^u) 
+ x^m \gamma_m\gamma_u \left(
H^{-3/2} \partial_u H  ( \tfrac{1}{12} -\tfrac{1}{8} \gamma_{013456} ) \epsilon_- 
+\tfrac12 H^{-1/6}\gamma_u \eta_-
\right) 
\\ & \qquad
+ x^I \gamma_I \gamma_u \left(
H^{-1} \partial_u H   (- \tfrac{1}{6} + \tfrac{1}{8} \gamma_{013456} ) \epsilon_- 
+\tfrac12 H^{1/3} \gamma_u\eta_-
\right) \,.
\end{split}
\label{DetermineEpsP}
\ee
The remaining equations \eqref{KSepsplusalpha} and \eqref{KSepsplus2} then determine $\rho_{+A}$.
The condition $\gamma^A \rho_{+A} = 0$ implies 
\be
H^{1/2} \gamma^\alpha \partial_\alpha \epsilon_+ + \gamma_u \partial_u \epsilon_+ 
+ H^{-1} \partial_u H \gamma_u  (- \tfrac{1}{6} + \tfrac{1}{4} \gamma_{013456} )  \epsilon_+ = 0 \,.
\label{DetermineDetermine} 
\ee
Substituting \eqref{DetermineEpsP} into \eqref{DetermineDetermine} leads to a first-order differential equation $\tilde \epsilon_+(x^\alpha,u)$ and second-order differential equations for $\epsilon_-$.
These have to be solved together with \eqref{DetermineEpsM}.

\vspace{1em}
\noindent \emph{An AdS-style solution.}
Instead of directly attacking these equations, I instead choose to start with an intelligent choice of $\eta_-$ which will lead to solutions.
After some trial, error and recollection of the structure of AdS spinors discussed in section \ref{sec:solutions}, I take:
\be
\eta_- = \tfrac{1}{12} H^{-4/3} \partial_u H \gamma_u \gamma_{013456} \epsilon_- \,.
\label{clevereta}
\ee
This implies that the equations \eqref{KSepsminusalpha} and \eqref{KSepsminus2} (or equivalently \eqref{DetermineEta} and \eqref{DetermineEpsM}) become
\begin{subequations}\label{someclevereqns}
\begin{align}
0 & = \partial_\alpha \epsilon_- - \tfrac{1}{12} H^{-3/2} \partial_u H \gamma_\alpha \gamma_u ( 1 - \gamma_{013456} ) \epsilon_- \,,
\label{cleveralpha}
\\[4pt]
0 & = \partial_u \epsilon_- + \tfrac{1}{12} H^{-1} \partial_u H \gamma_{013456} \epsilon_- \,.
\label{clever2}
\end{align} 
\end{subequations}
These equations can be explicitly solved as follows.
First of all, the appearance of the projector $1-\gamma_{013456}$ ensures that $\partial_\alpha \partial_\beta \epsilon_- = 0$.
Then \eqref{cleveralpha} is solved by 
\be
\epsilon_- = \tilde \epsilon_-(u) + \tfrac{1}{12} H^{-3/2} \partial_u H x^\alpha \gamma_\alpha \gamma_u (1 - \gamma_{013456} ) \tilde \epsilon_-(u) \,.
\ee
Inserting this solution into \eqref{clever2}, and
assuming that $(1-\gamma_{013456})\tilde \epsilon_- \neq 0$, leads to a differential equation for $\tilde \epsilon_+$:
\be
0  = \partial_u \tilde \epsilon_- + \tfrac{1}{12} H^{-1} \partial_u H \gamma_{013456} \tilde \epsilon_- \,,
\ee
as well as one for the otherwise entirely unconstrained function $H$:
\be
%\partial_u^2 H  - \tfrac{4}{3} (\partial_u H)^2 H^{-1}= 0 \,.
\partial_u (H^{-4/3} \partial_u H ) = 0 \,.
\label{Hequation}
\ee
Rather elegantly, this singles out the near-horizon solution of the M5-brane with $H \sim u^{-3}$. (In fact the general solution to this equation is $H = ( A u + B)^{-3}$ for some constants $A,B$, but this can always be put into the near-horizon form by a simple change of coordinates.)
Assuming \eqref{Hequation} is satisfied, the solution for $\tilde \epsilon_-$ is easily found to be:
\be
\tilde \epsilon_- = \left( H^{-1/12} \tfrac12 ( 1 +  \gamma_{013456} ) + H^{1/12} \tfrac12 ( 1 -  \gamma_{013456} ) \right) \epsc_-\,,
\ee
where $\epsc_-$ is a constant spinor.
The full solution for $\epsilon_-$ is then: 
\be
\epsilon_- = 
 H^{-1/12} \tfrac12 ( 1 +  \gamma_{013456} ) \epsc_-
+ H^{1/12} \left[
1 + \tfrac{1}{6} H^{-3/2} \partial_u H  x^\alpha \gamma_\alpha \gamma_u 
\right]\tfrac12 ( 1 - \gamma_{013456} )\epsc_-\,,
\label{epsminusSOLUTION}
\ee
which can be noted to essentially take the form, up to the choice of coordinate and projector, of  a Killing spinor for AdS${}_3$, compare \eqref{ADSeps} and \cite{Lu:1998nu}.

Now I turn to the Killing spinor equations for $\epsilon_+$. 
With the choice \eqref{clevereta} of $\eta_-$, the general expression \eqref{DetermineEpsP} for $\epsilon_+$ becomes:
\be
\epsilon_+ = \tilde \epsilon_+ (x^\alpha, u) 
+ \tfrac{1}{12}  H^{-3/2} \partial_u H x^m  \gamma_m\gamma_u ( 1 - \gamma_{013456} ) \epsilon_- 
-\tfrac16 H^{-1} \partial_u H x^I \gamma_I \gamma_u (1- \gamma_{013456} ) \epsilon_- \,.
\ee
Substituting into the remaining equations \eqref{KSepsplusalpha} and \eqref{KSepsplus2}, the shift symmetry parameter $\rho_{+A}$ is determined to be 
\be
\begin{split}
\rho_{+ \alpha}& = \tfrac{1}{36} H^{-7/3} \partial_u H \gamma_\alpha x^I \gamma_I ( 1  - \gamma_{013456} ) \epsilon_-  %+\tilde \rho_{+\alpha}(x^\alpha, u) 
\,,\\[4pt]
\rho_{+u} &= \tfrac{1}{6} (- H^{-4/3} \partial_u^2 H +H^{-7/3} (\partial_u H)^2 ) \gamma_u x^I \gamma_I ( 1  - \gamma_{013456} ) \epsilon_-
%+ \tilde \rho_{+ u}(x^\alpha, u) 
 \,.
\end{split}
\label{cleverrho}
\ee
This satisfies $\gamma^A \rho_{+A} = 0$ as long as $H$ obeys \eqref{Hequation}.

Then one is left to solve for $\tilde \epsilon_+$. % and $\tilde \rho_{+A}$.
Here I make the assumption that $\rho_{+A}$ has been fully determined: I will find solutions involving further contributions to $\rho_{+A}$ below.
Then $\tilde \epsilon_+$ obeys the  equations \eqref{KSepsplusalpha} and \eqref{KSepsplus2} omitting the $\rho_{+A}$ terms.
These equations are identical to those for $\epsilon_-$, \eqref{cleveralpha} and \eqref{clever2}, so the structure of the solution \eqref{epsminusSOLUTION} can be reused.
This gives a solution for $\epsilon_+$ of the following form:
\be
\begin{split} 
\epsilon_+ & = 
 H^{-1/12} \tfrac12 ( 1 +  \gamma_{013456} ) \epsc_+
+ H^{1/12} \left[
1 + \tfrac{1}{6} H^{-3/2}  \partial_u H  x^\alpha \gamma_\alpha \gamma_u 
\right]\tfrac12 ( 1 - \gamma_{013456} )\epsc_+ 
\\[4pt] & \qquad
+ \tfrac{1}{12}  H^{-3/2} \partial_u H x^m  \gamma_m\gamma_u ( 1 - \gamma_{013456} ) \epsilon_- 
-\tfrac16 H^{-1} \partial_u H x^I \gamma_I \gamma_u (1- \gamma_{013456} ) \epsilon_- \,.
\end{split}
\label{epsplusSOLUTION}
\ee
Together $\epsilon_-$ and $\epsilon_+$ given by \eqref{epsminusSOLUTION} and \eqref{epsplusSOLUTION} provide 32 Killing spinors for the M2NC background, where $H$ must obey \eqref{Hequation}, and $\eta_-$ and $\rho_{+A}$ are given by \eqref{clevereta} and \eqref{cleverrho}.
If $H$ does not obey \eqref{Hequation}, a solution to the Killing spinor equations with 16 independent components can instead be found.
Tracing through the above derivation, this is given by:
\be
\begin{split}
\epsilon_\pm & =  H^{-1/12} \tfrac12 ( 1 +  \gamma_{013456} ) \epsc_\pm\,,\quad
\eta_- = \tfrac{1}{12} H^{-4/3} \partial_u H \gamma_u \gamma_{013456} \epsilon_- \,,\quad
\rho_{+A} = 0 \,.
\end{split}
\ee
So the supersymmetry enhancement of the M5 solution when going to its AdS${}_4\times$ S${}^7$ limit survives in the M2NC background obtained using the large $N$ scaling.

\vspace{1em}
\noindent \emph{A solution with infinite-dimensional supersymmetry enhancement.}
However, the above solution is not the end of the story.
I can also find solutions with $\epsilon_- = 0 = \eta_-$ and $\epsilon_+ = \epsilon_+(x^\alpha,u)$, $\rho_{+A} = \rho_{+A}(x^\alpha, u)$.
Such solutions obey the equation \eqref{DetermineDetermine}.
If I suppose that $\gamma^\alpha \partial_\alpha \epsilon_+ = 0$, then I can solve this as:
\be
\epsilon_+ = H^{-1/12} \tfrac12 ( 1 + \gamma_{013456} ) \theta_+ (x^\alpha) + H^{5/12} \tfrac12 ( 1 - \gamma_{013456} ) \theta_+ (x^\alpha) \,,\quad
\gamma^\alpha \partial_\alpha  \theta_+ = 0 \,,
\label{epsPlusSolnEnhance}
\ee
involving an arbitrary solution $\theta_+$ of the two-dimensional massless Dirac equation.
This gives infinitely many further solutions of the Killing spinor equations, which by linearity can be added to the 32 component solution found above (or the 16 component solution if $H$ is not of the near-horizon form). For $\theta_+$ constant, the first term of \eqref{epsPlusSolnEnhance} in fact recovers part of the solution already found, while the second term in \eqref{epsPlusSolnEnhance} then seems to give a further $8$ independent solutions. 
This supersymmetry enhancement does not care about the precise form of $H(u)$, and 
(again) relies on the presence of the extra fermionic shift symmetries.
A deeper understanding of this mechanism, and a full accounting of the number of Killing spinors for longitudinally-dependent solutions, requires further work.
I leave this for the future, and instead turn to analyse the smeared solution, which will still allow for an infinite-dimensional enhancement in however a more restrictive manner.

\subsection{Smeared}
\label{sec:M5smeared}

I next consider the limit of the M5 solution smeared on the longitudinal M2NC direction $X^2$.
This limit was previously taken in \cite{Lambert:2024ncn}.
The harmonic function of the smeared M5 is:
\be
H = 1 + \frac{\tilde L^2}{X^IX^I} \,,\quad \tilde L^2 \sim R_2^{-1} N \ell_p^3 \,,
\ee
where $R_2$ is the radius of the $X^2$ direction.
The prescription for the M2NC limit is then simply
\be
(X^\alpha, X^2) = c (x^\alpha, u) \,,\quad
(X^m, X^I) = c^{-1/2} (x^m, x^I) \,,
\ee
implying $R_2 = c r_u$,
under which the harmonic function is manifestly finite, and furthermore the electric field strength involves no subleading terms.
This therefore leads to an M2NC geometry with a non-trivial electric field strength but with $\lambda_{abcd} = 0$.
This background has the form:
\be
\begin{split} 
\tau^\alpha & = H^{-1/6} \dd x^\alpha \,,\quad \tau^u= H^{1/3} \dd u \,,\quad
e^m = H^{-1/6} \dd x^m \,,\quad e^I = H^{1/3} \dd x^I \,,\\[4pt]
f_{(4)} & = - \tfrac{1}{3!} \epsilon_{IJKL} \partial_L H \dd x^I \wedge \dd x^J \wedge \dd x^K \wedge \dd u \,,\quad 
\lambda_{abcd} = 0 \,,\\[4pt]
H & = 1 + \frac{\tilde \ell^2}{x^Ix^I}\,,\quad \tilde \ell^2 \sim  r_u^{-1} N \ell_p^3 \,.
\end{split} 
\ee
This leads to non-vanishing $f_{Aabc}$ and $T_{a}{}^{\{AB\}}$ with components 
\be
f_{u IJK} = H^{-4/3} \epsilon_{IJKL} \partial_L H \,,\quad
T_{I\alpha}{}^\beta = -\tfrac16 H^{-4/3} \partial_I H \,,\quad
T_{Iu}{}^u = \tfrac13 H^{-4/3} \partial_I H  \,.
\ee
so this can only be a solution of the half-maximal version of M2NC SUGRA.
The components of the connections \eqref{connections} are:
\begin{align} 
\omega_{\mu}{}^{ab} & : \quad \omega_{\hat m}{}^{I n} = \tfrac16 H^{-3/2}\partial^I H \delta_n^m \,,\quad \omega_{\hat I}{}^{JK} = \tfrac23 H^{-1} \delta_I^{[J} \partial^{K]} H \,,
\end{align} 
with $\omega_{\mu}{}^{AB}$, $\omega_{\mu}{}^{Aa}$ and $b_\mu$ all zero.
Then, plugging the above configuration into the equations of motion shows they are satisfied as long as $H$ is harmonic in the $X^I$ directions.

The Killing spinor equations are:
\begin{subequations}\label{KSansatzM5smeared}
\begin{align} 
\label{KS1M5}
0& =  \left(
\tfrac12  T_{a}{}^{\{AB\}} \gamma_B \gamma^a 
+\tfrac14 ( \eta^{AB} - \tfrac13 \gamma^A \gamma^B ) \tfrac16 f_{Bcde} \gamma^{cde} \right) \epsilon_+
\,,\\[4pt] 
0 & = \left(e^\mu{}_a \partial_\mu  + \tfrac{1}{4} \omega_a{}^{bc} \gamma_{bc}  
+\tfrac{1}{12} \tfrac{1}{6}  f_{A bcd} ( \gamma_a{}^{bcd} - 6 \delta_a^b \gamma^{cd} ) \gamma^A \right) \epsilon_+  \,,
\label{KS2M5}
\\[4pt]
0 & = \tau^{\mu}{}_A \partial_\mu \epsilon_+
+ \rho_{+A}  \,.
\label{KS3M5}
\end{align}
\end{subequations}
The first equation here, \eqref{KS1M5}, again boils down to a worldvolume projection condition:
\be
\epsilon_+ = \gamma_{013456} \epsilon_+ \,.
\label{M5proj}
\ee
Using this, or the equivalent form $\gamma_I \epsilon_+ = \tfrac16 \epsilon_{KJLI} \gamma^I \gamma^u \epsilon_+$, repeatedly in \eqref{KS2M5} one finds:
\be
\partial_m \epsilon_+ = 0 \,,\quad
\partial_I \epsilon_+  + \tfrac{1}{12} \partial_I \ln H \epsilon_+ = 0 \,,
\ee
solved by $\epsilon_+ = H^{-1/12}(x^I) \theta_+(x^\alpha,u)$, where $\theta_+$ is projected as in \eqref{M5proj}.
Finally, the last equation \eqref{KS3M5} is indifferent to this projection, and leads trivially to a solution for $\rho_{+A}$. Then the final task is to solve $\gamma^A \rho_{+A} = 0$, which implies
\be
\gamma^A \tau^\mu{}_A \partial_\mu \epsilon_+ =  0 
\Rightarrow
\gamma^\alpha  \partial_\alpha \theta_+ + H^{-1/2} \gamma^u \partial_u \theta_+ 
= 0 \,.
\ee
Differentiating with $\partial_I$, it follows that each term must separately be zero:
\be
\partial_u \theta_+ = 0 \,,\quad 
\gamma^\alpha \partial_\alpha \theta_+ =0 \,.
\ee
So $\theta_+$ is independent of $u$, and solves the two-dimensional massless Dirac equation. 
This leads to a solution
\be
\epsilon_+ = H^{-1/12}(x^I) \theta_+(x^\alpha) \,,\quad \theta_+(x^\alpha) = \gamma_{013456} \theta_+(x^\alpha)\,,
\label{halfsolution}
\ee
with $\theta_+(x^\alpha)$ an arbitrary solution of the ($1+1$)-dimensional massless Dirac equation.
A general solution of this equation, using \eqref{thetasolnlc}, can be written as:
\be
\theta_+(x^\alpha) = \theta_+^{(0)}+ (\gamma^0 + \gamma^1) \chi_+(x^0 + x^1) + (\gamma^0 - \gamma^1) \tilde \chi_+(x^0 - x^1) \,,
\label{thetasoln}
\ee
where $\theta_+^{(0)}$ is constant, and the spinors $\chi_+(x^0+x^1)$ and $\tilde \chi_+(x^0-x^1)$ obey
\be
\chi_+ = - \gamma_{013456} \chi_+ \,,\quad
\chi_+ = + \gamma_{012} \chi_+ \,,\quad
\tilde \chi_+ = - \gamma_{013456} \tilde \chi_+ \,,\quad
\tilde \chi_+ = + \gamma_{012} \tilde \chi_+ \,,\quad
\ee
and further are arbitrary functions of the lightcone combinations $x^0\pm x^1$.
The Killing spinor \eqref{halfsolution} takes the form of the usual M5 Killing spinor with the additional projection $\epsilon \rightarrow \epsilon_+$ and the enhancement of $\theta_+$ to include the $x^\alpha$-dependent terms in \eqref{thetasoln}.
So again there is an infinite-dimensional enhancement of supersymmetry.
Note that there is no other enhancement of supersymmetry on replacing $H$ with its near-horizon form.

It is worthwhile at this point to attempt to connect to the field theoretic results of \cite{Lambert:2024yjk, Lambert:2024ncn}.
To do so, recall that the M2NC limit of the M5 solution is arranged in the form:
\be
\begin{tabular}{c|ccccccccccc}
 & 0 & 1 & 2 & 3 & 4 & 5 & 6 & 7 & 8 & 9 & 10 \\\hline 
\text{M2NC} & $\times$  & $\times$   & $\times$  & -- & -- & -- & -- & -- & -- & -- & -- \\[2pt]
\text{M5}& $\times$  & $\times$   & --  & $\times$ & $\times$& $\times$& $\times$& -- & -- & -- & --
\end{tabular}
\label{M2M5}
\ee
and the enhanced Killing spinor solution of the smeared M5 solution depends on the coordinates $x^\alpha=(x^0,x^1)$ and $x^I = (x^7,x^8,x^9,x^{10})$.
Dimensional reduction on one of the other M5 worldvolume directions, say $x^6$, leads to a type IIA D2NC limit:
\be
\begin{tabular}{c|cccccccccc}
 & 0 & 1 & 2 & 3 & 4 & 5 & 7 & 8 & 9 & 10 \\\hline 
\text{D2NC} & $\times$  & $\times$   & $\times$  & -- & --  & --  & -- & -- & -- & --\\[2pt]
\text{D4}& $\times$  & $\times$   & --  & $\times$ & $\times$& $\times$ & -- & -- & -- & --
\end{tabular}
\label{D2D4}
\ee
T-dualising this on a further worldvolume direction, say $x^3$, gives a type IIB D3NC limit of a smeared D3 brane:
\be
\begin{tabular}{c|cccccccccc}
 & 0 & 1 & 2 & 3 & 4 & 5 & 7 & 8 & 9 & 10 \\\hline 
\text{D3NC} & $\times$  & $\times$   & $\times$  & $\times$ & --  & -- & -- & -- & -- & -- \\[2pt]
\text{D3}& $\times$  & $\times$   & --  & -- & $\times$& $\times$ & -- & -- & -- & --
\end{tabular}
\label{D3D3}
\ee
These manipulations have only reduced or dualised on the directions $x^m$, on which the background and Killing spinors do not depend.
One can conclude that the D3NC limit of the smeared D3 brane should share the supersymmetries of the original M2NC--M5 case: that is, there are 8 `conventional' supersymmetries (corresponding to $\theta_+$ constant in \eqref{halfsolution}) plus an infinite-dimensional enhancement peculiar to the non-relativistic setting (corresponding to $\theta_+$ arbitrary solutions of the (1+1)-dimensional massless Dirac equation).
In \cite{Lambert:2024yjk} the supersymmetries of the proposed field theory dual of the configuration \eqref{D3D3} were analysed.
Half the supersymmetries in the field theory were observed to become local symmetries, and it was argued these would not be present in the geometric description. 
It was then conjectured that the underlying type IIB non-relativistic SUGRA based on the D3NC limit should in turn only have half the usual amount of supersymmetry.
This matches the fact here that the smeared solution is only present in the half-maximally supersymmetric version of M2NC SUGRA. 
In addition, the field theory enjoyed an enhanced superconformal symmetry. This enhancement would seem to be mirrored on the supergravity side by the appearance of the left- and right-moving Killing spinors solving the massless Dirac equation in $(1+1)$-dimensions.
Similar features were observed in \cite{Lambert:2024ncn} on studying the M5NC limit of the M2 brane, which is argued there to be equivalent to the `reciprocal' M2NC limit of the M5 brane, which is our starting point \eqref{M2M5}.

It would be good to make these connections more precise. A small further step one could imagine taking would be the following. In \cite{Lambert:2024ncn} the bosonic Killing symmetries of the smeared non-Lorentzian brane solutions involved were also analysed. One might imagine verifying from closure of the supersymmetry algebra that the supersymmetry transformations associated with the Killing spinors square to the expected bosonic symmetries (which also involve a conformal enhancement leading to the appearance of Virasoro symmetry). However, for the smeared solutions in the half-maximal version of M2NC SUGRA this is simply impossible as there the supersymmetry algebra does not close into diffeomorphisms \cite{Bergshoeff:2024nin}.

A different T-duality of \eqref{D2D4}, this time on the longitudinal spatial direction $x^1$, would map to a D1NC limit of a D3: 
\be
\begin{tabular}{c|cccccccccc}
 & 0 & 1 & 2 & 3 & 4 & 5 & 7 & 8 & 9 & 10 \\\hline 
\text{D1NC} & $\times$  & --   & $\times$  & -- & --  & --  & -- & -- & -- \\[2pt]
\text{D3}& $\times$  & --   & --  & $\times$ & $\times$& $\times$ & -- & -- & -- & --
\end{tabular}
\label{D1D3}
\ee
While the initial M5 background is independent of $x^1$ (due to the smearing), the Killing spinor solution \eqref{halfsolution} is not.
Therefore the Killing spinors with the $x^\alpha$ dependence (which necessarily involves both $x^0$ and $x^1$ simultaneously) do not survive the T-dualisation process.
A further T-duality on a worldvolume direction leads to a D2NC limit of a D2 which lifts to the M2NC limit of an M2 from section \ref{sec:M2} (equivalently, one could directly U-dualise \eqref{M2M5} on $x^1, x^5,x^6$ to obtain this M2NC/M2 configuration).
Accordingly, the enhancement of supersymmetry is not present in the supergravity description of \eqref{D1D3} or of the M2NC limit of the smeared M2 (though it may survive non-locally).
This seems in line with what was found for the smeared M2 in section \ref{sec:M2smeared}, and with the field theory symmetries of the D1NC limit of the D3 also discussed in \cite{Lambert:2024yjk}.

Notice that it is difficult to make the above arguments with respect to the scaled solutions.

\subsection{Spherical and scaled}
\label{sec:M5spherical}

Another variant of an M2NC limit of the M5 solution can be constructed by first switching to spherical coordinates in the transverse space, and then choosing the radial coordinate to correspond to an M2NC longitudinal coordinate. This is possible because for the M2NC limit of the M5 solution, there is a single direction which is longitudinal in the M2NC geometry and transverse to the brane. In the M2NC limit of the M2 brane, on the other hand, there are two such directions, and it would be less natural (though not necessarily impossible) to take the radial direction to be one of these.

The reason for considering this more exotic limit is that it should be related, by dimensional reduction and T-duality, to the limit of the D3 brane solution used in \cite{Fontanella:2024kyl,Fontanella:2024rvn} to propose a version of \adsS{5}{5} holography in the context of non-relativistic string theory (NRST), building on the earlier approach of \cite{Gomis:2005pg}.
NRST is based on a stringy Newton-Cartan (SNC) geometry, and so only ever has one spatial longitudinal direction, which in \cite{Fontanella:2024kyl,Fontanella:2024rvn} is taken to be the radial coordinate of the D3 brane solution.

The reduction and duality connecting these two set-ups is as follows: the M2NC limit of the M5 \eqref{M2M5} reduces on the common longitudinal direction $x^1$ to an SNC limit of the D4 solution:
\be
\begin{tabular}{c|cccccccccc}
 & 0  & 2 & 3 & 4 & 5 & 6 & 7 & 8 & 9 & 10 \\\hline 
\text{SNC} & $\times$  & $\times$  & -- & --  & -- & --  & --  & -- & -- & -- \\[2pt]
\text{D4}& $\times$  & --   & $\times$  & $\times$ & $\times$& $\times$ & -- & -- & -- & --
\end{tabular}
\label{F1D4}
\ee
and T-dualising on a D4 worldvolume direction gives an SNC limit of the D3 brane. Alternatively, one could S-dualise the D1NC/D3 configuration \eqref{D1D3}.
In the absence of the full formulation of non-relativistic type IIB supergravity (for the bosonic sector, see \cite{Bergshoeff:2023ogz}), I focus on the analogous M2NC M5 example with a view to gaining some insights into the supersymmetric properties of non-Lorentzian geometries obtained by this alternative limit.

In spherical coordinates, the original M5 configuration is:
\be
\label{M5}
\begin{split} 
\dd s^2  & = H^{-1/3} ( - (\dd X^0)^2 + (\dd X^1)^2 + \dd X^m \dd X^m ) 
+ H^{2/3} ( \dd R^2 + R^2 \dd \Omega_4^2)  \,,\\[4pt]
F_{(4)} & = \mathrm{Vol}(S^4) R^4 \frac{\partial H}{\partial R}\,,
\end{split}
\ee
where as before $m=3,4,5,6$ and $H = 1 + \left( \frac{L}{R} \right)^3$, $L^3 \sim N \ell_p^3$.
The unit metric on the four-sphere can be given in terms of embedding coordinates $Y^\Sigma$, $\Sigma=1,\dots,5$, obeying $Y^\Sigma Y^\Sigma = 1$, as $\dd \Omega_4^2 = \dd Y^\Sigma \dd Y^\Sigma$.
I now apply the M2NC limit, incorporating the extra scaling of the constant in the harmonic function, in the form:
\be
(X^0,X^1,R) = c (x^0,x^1,r) \,,\quad
X^m = c^{-1/2} X^m \,,\quad L^3 = c^3 \ell^3 \,,
\ee
which gives a finite $H = 1 + \ell^3/r^3$.
However, the spherical directions in the metric do not have the desired $c$-dependence to be transverse directions in an M2NC limit.
One needs to engineer an additional limit of the sphere.
A quick and dirty way to do this -- similar to \cite{Fontanella:2024kyl,Fontanella:2024rvn} --  would be something like the following.
Write the metric on the sphere as:
\be
\dd \Omega_4^2 = \frac{ \delta_{IJ} - \delta_{IJ} Y^K Y^K + Y_I Y_J}{1-Y^K Y^K} \dd Y^I \dd Y^J 
\ee
having let $Y^\Sigma=(Y^I, Y^5)$ with $Y^I Y^I + Y^5 Y^5 = 1$.
Then let $Y^I = c^{-3/2} x^I$ so
\be
\dd \Omega_4^2 = c^{-3} \dd x^I \dd x^I+ O(c^{-6}) 
\,,\quad
F_{(4)} = c^{-3} \tfrac{1}{4!} \epsilon_{IJKL} \dd x^I \wedge \dots \wedge \dd x^L r^4 \frac{\partial H}{\partial r} \,.
\ee
This could perhaps be viewed as `zooming in' on the sphere in an infinitesimal vicinity of the north or south pole. However I leave further decipherment of the physical and geometric meaning of this limit aside.
I focus simply on studying the properties of the M2NC background resulting from the above procedure, which has the form:
\be
\begin{split}
\tau^\alpha &= H^{-1/6} \dd x^\alpha \,,\quad
\tau^r = H^{1/3} \dd r \,,\quad
e^m = H^{-1/6} \dd x^m \,,\quad
e^I = r H^{1/3} \dd x^I \,,
\\[4pt]
c_{\mu\nu\rho} &= 0 \,,\quad
\lambda_{789 10} = \tfrac12 H^{-4/3} \frac{\partial H}{\partial r}  = - \lambda_{3456} \,.
\end{split}
\ee
Again, for $H$ an arbitrary function of $r$, this obeys the constraints of the maximally supersymmetric version of M2NC, as well as the Poisson equation.
Does it admit Killing spinors?
I cut to the chase and focus on the near-horizon version of the solution, where $H = r^{-3}$ (I set the constant $\ell=1$ for simplicity), for which $e^I = \dd x^I$, and the connections \eqref{connections} simplify to:
\begin{subequations}
\begin{align}
 \omega_\mu{}^{AB} :\quad &  \omega_{\hat \alpha}{}^{\beta r} = \tfrac12 r^{1/2} \delta_\alpha{}^\beta \,,\quad
   \omega_\mu{}^{Aa} :\quad   \omega_{\hat m}{}^{r n} = - \tfrac12 r^{1/2}  \delta_m^n \,,\quad   \omega_\mu{}^{ab}  =  0 \,,\quad b_{\mu} = 0 \,.
\end{align}
\end{subequations}
Relative to the scaled M5 limit of section \ref{sec:M5scaled}, the only difference here is that the components $\omega_{\hat I}{}^{r J}$ of the boost connection are now zero (the flat index $r$ here is identified with the flat index $u$ there, both being the original `$A=2$' longitudinal index).
This means that one obtains exactly the same Killing spinor equations as in \eqref{KSM5Scaled}, replacing there $\partial_u \rightarrow \partial_r$ and $H\rightarrow r^{-3}$, except for that involving $\partial_I \epsilon_+$, for which the boost connection contribution drops out and, taking into account the different form of $e^I$, one instead has:
\be
0  = \partial_I \epsilon_+  - \tfrac{3}{8} \gamma_r \gamma_I \gamma_{013456} \epsilon_- 
-\tfrac12 \gamma_I \eta_- \,.\label{KSepsplusIr}
\ee
I can first find an AdS-type solution.
To do so, consistent with the analysis of section \ref{sec:M5scaled}, I choose $\eta_- = - \tfrac14 \gamma_r \gamma_{013456} \epsilon_-$ so that the system of equations that must be (re)solved can be written:
\begin{subequations}
\label{KSM5Polar}
\begin{align}
\partial_\alpha \epsilon_{\pm} &	= - \tfrac{1}{4} r^{1/2} \gamma_\alpha \gamma_r ( 1 -  \gamma_{013456}) \epsilon_{\pm} \,,
\qquad 
\partial_r \epsilon_{\pm} = \tfrac{1}{4r} \gamma_{013456} \epsilon_{\pm} \,,\\[4pt]
\partial_m \epsilon_+ & = \tfrac14 r^{1/2} \gamma_r \gamma_m ( 1 -  \gamma_{013456}) \epsilon_-\,, \quad 
\partial_I \epsilon_+  = \tfrac12 \gamma_r \gamma_I \gamma_{013456} \epsilon_- \,,
\end{align} 
\end{subequations}
together with $\partial_m \epsilon_- = 0$, $\partial_I \epsilon_- = 0$, and where I have pre-emptively set $\rho_{+A} = 0$.
Previously, $\rho_{+A}$ had to be non-zero in order to absorb contributions to $\epsilon_+$ that were linear in $x^I$.
This time around, this is not needed.
It can be checked that the equations \eqref{KSM5Polar} are exactly solved by 
\begin{align}
\epsilon_-& = r^{1/4} \tfrac12 ( 1 +  \gamma_{013456}) \epsilon^{(0)}_-
+ ( r^{-1/4} + \tfrac12 r^{1/4} x^\alpha \gamma_r \gamma_\alpha )  \tfrac12 ( 1 - \gamma_{013456}) \epsilon^{(0)}_-\,,\\[4pt]
\notag
\epsilon_+& = r^{1/4} \tfrac12 ( 1 +  \gamma_{013456}) \epsilon^{(0)}_+
+ ( r^{-1/4} + \tfrac12 r^{1/4} x^\alpha \gamma_r \gamma_\alpha )  \tfrac12 ( 1 - \gamma_{013456}) \epsilon^{(0)}_+
\\[4pt] & \qquad + \tfrac14 x^m r^{1/2} \gamma_r \gamma_m ( 1 - \gamma_{013456})  \epsilon_- 
+ \tfrac12 x^I \gamma_r \gamma_I \gamma_{013456} \epsilon_- \,.
\end{align} 
Thus the solution obtained by treating the M5 radial coordinate as longitudinal, and engineering a flat space limit of the spherical part of the geometry, also gives a 32 component AdS-style Killing spinor solution of the maximally supersymmetric version of the M2NC SUGRA.
Again though, there is the possibility to find additional solutions of the form \eqref{epsPlusSolnEnhance}, which here is:
\be
\epsilon_+ = r^{1/4} \tfrac12 ( 1 + \gamma_{013456} ) \theta_+ (x^\alpha) + r^{-5/4} \tfrac12 ( 1 - \gamma_{013456} ) \theta_+ (x^\alpha) \,,\quad
\gamma^\alpha \partial_\alpha  \theta_+ = 0 \,.
\label{epsPlusSolnEnhancer}
\ee
Thus this alternative M2NC limit of the M5 shares the supersymmetric properties, including infinite-dimensional enhancement, of the limit with scaling discussed in section \ref{sec:M5scaled}.
Note that for $\theta_+$ non-constant, the Killing spinors \eqref{epsPlusSolnEnhancer} depend on $x^1$ and will not survive the dimensional reduction process leading to the type IIB picture that motivated taking the M2NC limit of the M5 in this way.

\section{Summary} 
\label{sec:summary}

I have discussed how for both the M2 and M5 solutions of 11-dimensional supergravity, the M2NC limit can be taken in two ways, and verified that both these methods indeed lead to supersymmetric solutions of the M2NC SUGRA.
The possibilities that I have considered can be summarised as follows: 

\begin{enumerate}
\item \emph{Scaled solutions.}
Here an extra scaling (of the number of branes) is introduced by hand when taking the limit of the original supergravity solution.
This leads to a solution of the M2NC SUGRA which is localised in longitudinal directions of the M2NC geometry.
This solution obeys the constraints of the maximally supersymmetric version of the M2NC SUGRA, and trivially obey the equations of motion, \emph{i.e.} the longitudinal coordinate dependence of the solutions is not determined by the equations of motion. 
These solutions are always supersymmetric, and there are additional Killing spinor solutions which only appear if the function $H$ characterising the background obeys certain conditions, which in particular accommodate the usual near-horizon limit.
In the M5 case, I can always find an infinite number of solutions to the Killing spinor equation, while in the M2 case this is only possible in the near-horizon limit. 

I also showed that the M2NC limit of the M5 solution allows for an alternative limit of this type, in which the spherical radial coordinate is taken to be a longitudinal M2NC direction.
This also gives rise to a supersymmetric solution of the maximally supersymmetric version of M2NC SUGRA.	 	

\item \emph{Smeared solutions.} Here the branes are smeared in the longitudinal directions of the M2NC geometry, leading to solutions localised only in transverse directions.
These solutions only satisfy the constraints of the half-maximally supersymmetric version of the M2NC SUGRA, and the equations of motion impose that the function $H$ appearing in the brane solutions is harmonic in the directions transverse both to the brane and in the M2NC geometry.
These solutions are supersymmetric, but there is no enhancement in the near-horizon limit.
However, for the M5 case an (infinite-dimensional) supersymmetry enhancement is always present.
The structure here is consistent with thinking of the M2NC limit itself, in a sense, as a 1/2 BPS object.
It is also compatible with the comments in \cite{Lambert:2024yjk} that suggested, from the field theory side, that half the supersymmetries of the bulk theory should be absent, and seems to match the  superconformal symmetry enhancement observed there.
\end{enumerate}

The time-honoured procedures of dimensional reduction and dualisation, already present in sections \ref{sec:M5smeared} and \ref{sec:M5spherical}, can be used to convert the M2 and M5 solutions of this paper into many, if not all, of the non-relativistic brane geometries produced in \cite{Lambert:2024uue, Lambert:2024yjk,Fontanella:2024rvn,Fontanella:2024kyl,Blair:2024aqz,Lambert:2024ncn,Harmark:2025ikv}.\footnote{\,``Exercise for the reader.''}
Simultaneously, one expects these procedures to map the underlying non-relativistic supergravity theories and their constraints.
Accordingly, I expect the behaviour exhibited in this paper is likely generic amongst the various known would-be solutions of non-Lorentzian supergravity obtained via similar string or brane non-relativistic limits.
Then longitudinally localised solutions would be compatible with maximally supersymmetric versions of non-relativistic supergravity, while transversely localised solutions would be compatible with half-maximally supersymmetric versions.
If the transversely localised solutions from smearing are indeed the correct backgrounds appearing in proposed non-relativistic AdS/CFT examples this implies that the unconventional half-maximal version of non-relativistic maximal supergravity proposed in \cite{Bergshoeff:2024nin} should be taken more seriously.
To support these claims, the supersymmetric completions of the multitude of type II string and brane non-relativistic limits should first be constructed (see \cite{Blair:2021waq,Bergshoeff:2023ogz,Lambert:2024ncn} for various bosonic sectors).

To choose the simplest example, the M2NC limits of the smeared M2 and M5 can be reduced on a longitudinal spatial direction to give Stringy Newton-Cartan (SNC) smeared D2 and D4 brane solutions, constructed using null T-duality in \cite{Harmark:2025ikv}. 
It follows automatically that these D-brane solutions obey the 10-dimensional constraints following from dimensional reduction of \eqref{halfcons}, which should appear in a half-maximal version of the full type IIA SNC maximal SUGRA descending from the 11-dimensional M2NC SUGRA.
This would then answer affirmatively the question raised in \cite{Harmark:2025ikv} concerning whether the solutions obtained there satisfy the constraints expected to follow from supersymmetry.

In general, as the smeared solutions do not depend on the longitudinal directions of the non-relativistic limit, it is always possible to dualise on these directions. This can be used to convert the limit to a null compactification or DLCQ (discrete lightcone quantisation) of a dual brane solution.\footnote{Thanks to Niels Obers and Ziqi Yan for a discussion highlighting this point.}
The relationship to DLCQ is central to the duality web of BPS decoupling limits \cite{Blair:2023noj, Gomis:2023eav, Blair:2024aqz,Harmark:2025ikv}.

For example, consider the M2 solution in the form \eqref{M2platonic}, assuming the harmonic function is smeared in the $\alpha=1,2$ directions. 
Then U-dualising on the $123$ directions leads to the following M2 solution:
\be
\begin{split}
\dd s^2 & = H^{-2/3} ( \tfrac{1}{c^3} (\dd x^3)^2 + 2 \,\dd x^0 \dd x^3 + (\dd x^4)^2 ) + H^{-1/3} ( \dd x^\alpha \dd x^\alpha + \dd x^I \dd x^I ) \,,\\[4pt]
C_{(3)} & = H^{-1} \dd x^0 \wedge \dd x^3 \wedge \dd x^4 \,,
\end{split}
\ee 
where $I=5,\dots,10$, and $H=H(x^I)$. The limit $c \rightarrow \infty$ now defines a null compactification of the $x^3$ direction.
On the other hand, when the solution is localised in the longitudinal directions, and independent of the transverse ones, this duality is not possible. 
It is tempting to speculate that this is a relevant observation for understanding the significance of the scaled solutions.
Note that the longitudinal dualities mapping to a DLCQ picture would also generically break any supersymmetries associated with Killing spinors depending on coordinates being dualised: the infinite-dimensional enhancement appearing for the M5 brane solutions carried such dependence.

An additional source (literally) of clarification regarding these solutions would be to include a brane source term, so as to view the brane solutions as solving the equations of motion following from the coupled action $S_{\text{SUGRA}} + N S_{\text{worldvolume}}$. 
The expectation would be that the brane solutions of non-relativistic supergravity are sourced by the corresponding worldvolume actions obtained by taking the target space non-relativistic limit in the worldvolume theory.
It seems hard to make sense of an additional rescaling of $N$ in this set-up, but further work is required to make a precise statement of the prospective issues.

For instance, consider the M2 configuration of section \ref{sec:M2}.
This should be sourced by an M2 worldvolume action describing an M2 that is \emph{not} longitudinally aligned with the longitudinal directions of the M2NC limit, which is what is often (implicitly or explicitly) assumed when deriving the M2 worldvolume action in this limit (various versions of M2 and M5 worldvolume actions in non-relativistic limits can be found in, for example, \cite{Garcia:2002fa,Kluson:2019uza,Ebert:2021mfu,Ebert:2023hba}).
Dimensionally reducing the M2NC limit of the M2 on a worldvolume direction turns this configuration into a string solution in the D2 Newton-Cartan limit, which leads to Matrix $2$-brane Theory (M2T) in the language of \cite{Gomis:2023eav, Blair:2023noj, Blair:2024aqz}. The string worldsheet action in this limit is constructed in \cite{Gomis:2023eav} and should source the corresponding supergravity solution: the limit of the supergravity theory in this case follows by dimensional reduction from 11 dimensions \cite{Blair:2021waq}, so the coupled equations of motion could be immediately checked in this setting. I leave this for future work.

Finally, I comment on the solutions discussed in section \ref{sec:solutions}.
Here a simple ansatz -- inspired by a similarly `innocuous' one in the context of the bosonic M5NC SUGRA \cite{Bergshoeff:2025grj} -- showed that arbitrary longitudinal three-geometries with never positive Ricci scalar are solutions of the maximally supersymmetric version of M2NC SUGRA.
In particular any local AdS${}_3$ geometry, such as the BTZ black hole, can appear.
Similarly in \cite{Bergshoeff:2025grj}, it was noted that a longitudinal six-dimensional Schwarzschild black hole is a solution of the M5NC SUGRA. These black hole solutions appear in the longitudinal part of the geometry, which retains a Lorentzian structure, which explains why they may still show up in a non-Lorentzian geometry. It would be nice to understand their properties further, including whether they can be obtained via limits or (null) duality from relativistic solutions.

\section*{Acknowledgements}

I am supported through the grants CEX2020-001007-S and PID2021-123017NB-I00, funded by MCIN/AEI/10.13039/501100011033 and by ERDF A way of making Europe.
This work benefitted, mostly long before it was started, from useful discussion and illuminating correspondence with: Eric Bergshoeff, Juan Nieto García, Troels Harmark, Johannes Lahnsteiner, Neil Lambert, Niels Obers, Jan Rosseel, Joseph Smith, Ziqi Yan. Further thanks to Neil Lambert, Niels Obers, Joseph Smith and Ziqi Yan for comments on the draft.

\addcontentsline{toc}{section}{References}

\bibliography{ChrisBib}

\end{document}